Standardized compliance matrixes for general anisotropic materials and a simple measure of anisotropy degree based on shear-extension coupling coefficient

Jiamin Zhao[1,2], Xiaoxiong Song[1], Bin Liu[1,2,*]

[1] *AML, Department of Engineering Mechanics, Tsinghua University, Beijing 100084, China*

[2]*CNMM, Department of Engineering Mechanics, Tsinghua University, Beijing 100084, China*

*Corresponding author: Tel.: 86-10-62786194; fax: 86-10-62781824.

*E-mail address*: liubin@tsinghua.edu.cn (Bin Liu).

**Abstract**

The compliance matrix for a general anisotropic material is usually expressed in an arbitrarily chosen coordinate system, which brings some confusion or inconvenience in identifying independent elastic material constants and comparing elastic properties between different materials. In this paper, a unique stiffest orientation based standardized (STF-OS) compliance matrix is established, and 18 independent elastic material constants are clearly shown. During the searching process for stiffest orientation, it is interesting to find from our theoretical analysis and an example that a material with isotropic tensile stiffness does not definitely possess isotropic elasticity. Therefore the ratio between the maximum and minimum tensile stiffnesses is not a correct measure of anisotropy degree. Alternatively, a simple and correct measure of anisotropy degree based on the maximum shear-extension coupling coefficient in all orientations is proposed.

## 1. Introduction

The generalized Hooke's law relating stresses to strains for anisotropic elastic materials can be written as

$$\begin{bmatrix}\varepsilon_{11}\\ \varepsilon_{22}\\ \varepsilon_{33}\\ \gamma_{23}\\ \gamma_{13}\\ \gamma_{12}\end{bmatrix}=\begin{bmatrix}\varepsilon_{11}\\ \varepsilon_{22}\\ \varepsilon_{33}\\ 2\varepsilon_{23}\\ 2\varepsilon_{13}\\ 2\varepsilon_{12}\end{bmatrix}=\begin{bmatrix}S_{11} & S_{12} & S_{13} & S_{14} & S_{15} & S_{16}\\ S_{21} & S_{22} & S_{23} & S_{24} & S_{25} & S_{26}\\ S_{31} & S_{32} & S_{33} & S_{34} & S_{35} & S_{36}\\ S_{41} & S_{42} & S_{43} & S_{44} & S_{45} & S_{46}\\ S_{51} & S_{52} & S_{53} & S_{54} & S_{55} & S_{56}\\ S_{61} & S_{62} & S_{63} & S_{64} & S_{65} & S_{66}\end{bmatrix}\begin{bmatrix}\sigma_{11}\\ \sigma_{22}\\ \sigma_{33}\\ \sigma_{23}\\ \sigma_{13}\\ \sigma_{12}\end{bmatrix}, \tag{1}$$

where $\sigma_{ij}$, $\varepsilon_{ij}$ and $S_{ij}$ are stress, strain and compliance matrix components, respectively. The compliance matrix $\mathbf{S}$ is a symmetric matrix, i.e. $S_{ij}=S_{ji}$. In many textbooks (Bower, 2011; Jones, 1998; Kaw, 2005; Ting and Horgan, 1996), it is written that a general anisotropic material has 21 independent elastic constants in a compliance matrix. If material possesses orientational symmetries, the number of independent elastic constants will reduce. For example, an orthotropic material, with three orthogonal planes of material symmetry, has only 9 independent elastic constants. In these textbooks, the compliance matrix of orthotropic material are written as

$$\mathbf{S}^{\mathbf{ortho}} = \begin{bmatrix} S_{11} & S_{12} & S_{13} & 0 & 0 & 0 \\ S_{12} & S_{22} & S_{23} & 0 & 0 & 0 \\ S_{13} & S_{23} & S_{33} & 0 & 0 & 0 \\ 0 & 0 & 0 & S_{44} & 0 & 0 \\ 0 & 0 & 0 & 0 & S_{55} & 0 \\ 0 & 0 & 0 & 0 & 0 & S_{66} \end{bmatrix} \tag{2}$$

When we observe these two compliance matrixes Eq.(1) and Eq.(2), we find that there is some inconsistence between them. The compliance matrix for orthotropic elastic material is usually expressed in a special coordinate system accounting for the intrinsic orientational symmetry, but the compliance matrix for general anisotropic elastic materials is expressed in an arbitrarily chosen coordinate system. The compliance matrix varies with different coordinate systems. If there is a standardized coordinate system, as in an orthotropic material, the compliance matrix becomes a standardized one, which provides convenience in comparing the elastic properties among different materials. This standardized compliance matrix can also clearly reveal the number of independent elastic material constants, e.g. 9 for orthotropic materials. However, to our best knowledge, no standardized compliance matrix for general anisotropic elastic materials has been proposed, which also brings students some confusion in the number of independent elastic material constants when they read textbooks (Bower, 2011; Jones, 1998; Kaw, 2005; Ting and Horgan, 1996).

The number of elastic constants has been a basic problem for a long time. The 21 independent elastic constants of anisotropic material are widely accepted after Green introduced the concept of strain energy and Load Kelvin proved the existence of Green’s strain energy function (Herakovich, 2012). However, Fedorov(Fedorov, 2012) showed that a definite choice of coordinate system may impose three conditions on elastic constants of anisotropic material, resulting in that the number of independent constants cannot exceed 18. Some other researchers also demonstrated the situation of 18 independent constants. Based on the theory of existence of longitudinal wave by Truesdell (Truesdell, 1966; Truesdell, 1968), Boulanger and Hayes (Boulanger and Hayes, 1995) discovered two components will vanish when the coordinate axis is the direction making young’s modulus reach an extremum value. The same phenomena occurred when the coordinate axis was set as the direction making the axial elastic coefficient extremum by Cowin and Mehrabadi (Cowin and Mehrabadi, 1995). They also found the third vanishing component through a specific rotation of coordinate system. Hence, 18 independent elastic constants are proved, which however has not been clearly presented to students through textbooks.

In this paper, we propose a unique standardized compliance matrix for general anisotropic materials to show the elastic properties in a unified way, such as independent elastic constants, to avoid confusion. The paper is structured as follows. In Section 2, we propose a unique stiffest orientation based standardized compliance matrix, and compare it with symmetry orientation based standardized compliance matrixes. We investigate the relation between the symmetry of tensile stiffness and the symmetry of the elastic properties in Section 3. By interestingly noting that the ratio between the maximum and minimum tensile stiffnesses is not a correct measure of anisotropy degree, we propose a simple and correct measure based on the maximum shear-extension coupling coefficient in Section 4. Conclusions are summarized in Section 5.

## 2. The definition and characteristics of a standardized compliance matrix for general anisotropic materials

### *2.1. The definition of the stiffest orientation based standardized compliance matrix*

To exclude the orientational arbitrariness of a coordinate system and obtain a unique standardized compliance matrix, a special coordinate system based on the material intrinsic orientation is needed. For orthotropic materials, the orientational symmetry can be adopted in choosing the coordinate system. For general anisotropic materials, however, there is no orientational symmetry. To deal with this situation, we propose that the material intrinsic orientation with extreme stiffness is used in establishing a standardized coordinate system $\hat{x}_1-\hat{x}_2-\hat{x}_3$, and the specific rules or conditions are suggested as follows.

**Condition I** The $\hat{x}_1$-axis makes $S_{11}$ reach the global minimum, i.e. the stiffest direction in the whole orientations;

**Condition II** The $\hat{x}_2$-axis makes $S_{22}$ global minimum within the plane perpendicular to $\hat{x}_1$-axis;

**Condition III** The $\hat{x}_3$-axis complies with the right-hand coordinate system rule;

**Condition IV** $S_{14}$ and $S_{25}$ reach their minimum values under above conditions.

The coordinate system $\hat{x}_1-\hat{x}_2-\hat{x}_3$ satisfying Condition I-IV is named as the stiffest orientation based standardized coordinate system, or STF-OS coordinate system for abbreviation. The corresponding compliance matrix in this coordinate system is named as **the stiffest orientation based standardized compliance matrix,** or **STF-OS compliance matrix** for abbreviation, and denoted as $\hat{\mathbf{S}}$. The first three conditions are obvious. For a general anisotropic material, as shown in Fig. 1, there are four coordinate systems satisfying Condition I-III, since one stiffest direction has two options for $\hat{x}_1$-axis, or $\hat{x}_2$-axis. The relations among these four candidate coordinate systems (a)-(d) are also shown in Fig. 1. Rotating Coordinate system (a) $\pi$ angle around $\hat{x}_1^a$-axis yields Coordinate system (b). Rotating Coordinate system (a) $\pi$ angle around $\hat{x}_2^a$-axis yields Coordinate system (c). Rotating Coordinate system (a) $\pi$ angle around $\hat{x}_3^a$-axis yields Coordinate system (d). The compliance matrixes under coordinate systems (a)-(d) are denoted as $\mathbf{S}^a$, $\mathbf{S}^b$, $\mathbf{S}^c$ and $\mathbf{S}^d$, respectively. Supposing

$$\mathbf{S}^a=\begin{bmatrix} S_{11}^a & S_{12}^a & S_{13}^a & S_{14}^a & S_{15}^a & S_{16}^a \\ & S_{22}^a & S_{23}^a & S_{24}^a & S_{25}^a & S_{26}^a \\ & & S_{33}^a & S_{34}^a & S_{35}^a & S_{36}^a \\ & (sym) & & S_{44}^a & S_{45}^a & S_{46}^a \\ & & & & S_{55}^a & S_{56}^a \\ & & & & & S_{66}^a \end{bmatrix} \tag{3a}$$

According to transformation relation among the four coordinate systems (see Appendix A), $\mathbf{S}^b$, $\mathbf{S}^c$ and $\mathbf{S}^d$ can be represented by the components of $\mathbf{S}^a$ as follows

$$\mathbf{S}^b = \begin{bmatrix} S_{11}^a & S_{12}^a & S_{13}^a & S_{14}^a & -S_{15}^a & -S_{16}^a \\ & S_{22}^a & S_{23}^a & S_{24}^a & -S_{25}^a & -S_{26}^a \\ & & S_{33}^a & S_{34}^a & -S_{35}^a & -S_{36}^a \\ & (sym) & & S_{44}^a & -S_{45}^a & -S_{46}^a \\ & & & & S_{55}^a & S_{56}^a \\ & & & & & S_{66}^a \end{bmatrix} \tag{3b}$$

$$\mathbf{S}^c = \begin{bmatrix} S_{11}^a & S_{12}^a & S_{13}^a & -S_{14}^a & S_{15}^a & -S_{16}^a \\ & S_{22}^a & S_{23}^a & -S_{24}^a & S_{25}^a & -S_{26}^a \\ & & S_{33}^a & -S_{34}^a & S_{35}^a & -S_{36}^a \\ & (sym) & & S_{44}^a & -S_{45}^a & S_{46}^a \\ & & & & S_{55}^a & -S_{56}^a \\ & & & & & S_{66}^a \end{bmatrix}$$

(3c)

$$\mathbf{S}^d = \begin{bmatrix} S_{11}^a & S_{12}^a & S_{13}^a & -S_{14}^a & -S_{15}^a & S_{16}^a \\ & S_{22}^a & S_{23}^a & -S_{24}^a & -S_{25}^a & S_{26}^a \\ & & S_{33}^a & -S_{34}^a & -S_{35}^a & S_{36}^a \\ & (sym) & & S_{44}^a & S_{45}^a & -S_{46}^a \\ & & & & S_{55}^a & -S_{56}^a \\ & & & & & S_{66}^a \end{bmatrix}$$

(3d)

Obviously, the compliance matrixes under these four coordinate systems are different. Therefore, we propose that the unique standardized compliance matrix is the one with minimal $S_{14}$ and $S_{25}$ among these four matrixes, i.e., Condition IV.

Condition I-IV can ensure the uniqueness of this standardized compliance matrix for a general anisotropic material. However, for some materials with special symmetries, the uniqueness of STF-OS compliance matrix may not be guaranteed by Condition I-IV. Therefore, a more general rule in determining a unique standardized compliance matrix is proposed as follows. Firstly, a specified sequence of compliance matrix components is introduced as

$(S_{11},S_{22},S_{33},S_{44},S_{55},S_{66},S_{12},S_{13},S_{14},S_{15},S_{16},S_{23},S_{24},S_{25},S_{26},S_{34},S_{35},S_{36},S_{45},S_{46},S_{56})$

and is schematically shown in Fig.2. Similar to Condition I-IV, we first find coordinate systems to make $S_{11}$ global minimum. For the rest components in the sequence, we then sequently search for their global minimums while keeping the previous components unchanged. It should be pointed out that for most situations, Condition I-IV are consistent with this general rule, and are relatively easier to use.

For any anisotropic material, adopting such unique STF-OS compliance matrix can eliminate the arbitrariness brought from different reference coordinate systems, and make it convenient to compare the elastic properties between different materials.

*2.2. The characteristic of the stiffest orientation based standardized compliance matrix for general anisotropic materials*

For general anisotropic materials, we will illustrate and prove that there are three zeroes in the upper triangle of the standardized compliance matrix, i.e. 18 independent elastic material constants.

Figure 3 shows an initially rectangular representative volume element (RVE) subject to a uniaxial tensile stress $\sigma_{11}$ along the $\hat{x}_1$-direction. In the following, we use proof by contradiction to demonstrate that the deformed RVE is still a rectangle. Firstly, we assume that the deformed RVE under $\sigma_{11}$ becomes a parallelogram as shown in Fig. 3(a), and we can note that the direction rotating anticlockwise slightly from $\hat{x}_1$-direction (denoted as a dotted line) is less stretchable and therefore has a larger tensile stiffness, which is contradictory to the condition that the global maximum stiffness is achieved in $\hat{x}_1$-direction. Hence the deformed RVE is still a rectangle as shown in Fig. 3(b), which implies that the $\sigma_{11}$ along the $\hat{x}_1$-direction cannot lead to the shear strain $\varepsilon_{12}$, i.e. $\hat{S}_{16}=0$. Similarly, $\hat{S}_{15}=0$ and $\hat{S}_{24}=0$ can be obtained. These conclusions can also be proved strictly in an analytical way.

Compliance components vary with the rotation of the coordinate system. When rotating the STF-OS coordinate system $\hat{x}_1-\hat{x}_2-\hat{x}_3$ around $\hat{x}_3$-axis through an angle $\theta_3$ (in this paper all rotations follow the right-hand rule), the compliance component $S_{11}$ can be expressed as

$$\begin{aligned} S_{11}(\hat{S}_{ij},\theta_3) = &\cos^4(\theta_3)\hat{S}_{11}+\sin^4(\theta_3)\hat{S}_{22}+\sin^2(\theta_3)\cos^2(\theta_3)\left(\hat{S}_{66}+2\hat{S}_{12}\right) \\ &+2\sin(\theta_3)\cos^3(\theta_3)\hat{S}_{16}+2\sin^3(\theta_3)\cos(\theta_3)\hat{S}_{26} \end{aligned} \tag{4}$$

According to Condition I and the above equation, $\hat{S}_{11}$ is the global minimum of $S_{11}$, and is achieved along $\hat{x}_1$-axis, i.e.,

$$\left.\frac{\partial S_{11}}{\partial \theta_3}\right|_{\theta_3=0}=2\hat{S}_{16}=0 \tag{5}$$

Similarly,

$$\left.\frac{\partial S_{11}}{\partial \theta_2}\right|_{\theta_2=0}=-2\hat{S}_{15}=0 \tag{6}$$

According to Condition II, $\hat{S}_{22}$ is the minimum of $S_{22}$ in the plane perpendicular to $\hat{x}_2$-axis, and is achieved along $\hat{x}_2$-axis, i.e.,

$$\left.\frac{\partial S_{22}}{\partial \theta_1}\right|_{\theta_1=0}=2\hat{S}_{24}=0 \tag{7}$$

Based on the above discussions, the final form of the STF-OS compliance matrix for general anisotropic materials is

$$\hat{\mathbf{S}}=\begin{bmatrix} \hat{S}_{11} & \hat{S}_{12} & \hat{S}_{13} & \hat{S}_{14} & 0 & 0 \\ & \hat{S}_{22} & \hat{S}_{23} & 0 & \hat{S}_{25} & \hat{S}_{26} \\ & & \hat{S}_{33} & \hat{S}_{34} & \hat{S}_{35} & \hat{S}_{36} \\ & (sym) & & \hat{S}_{44} & \hat{S}_{45} & \hat{S}_{46} \\ & & & & \hat{S}_{55} & \hat{S}_{56} \\ & & & & & \hat{S}_{66} \end{bmatrix} \tag{8}$$

It is clearly revealed that there are only 18 independent elastic material constants in the standardized compliance matrix, which looks different from Eq.(1) appearing in most textbooks. Here, we intentionally use “elastic material constants” to reflect their material intrinsic feature, which is independent of the coordinate orientation. Therefore, 18 independent elastic material constants together with three arbitrariness of the coordinate orientation result in 21 independent elastic constants in the usual form of the compliance matrix for general anisotropic materials. We suggest that including these statements in the textbooks will avoid possible confusions to readers.

In addition, it should be pointed out that for general anisotropic materials the inverse matrix of this standardized compliance matrix (Eq.(8)), i.e., the stiffness matrix, does not have three zeroes, which implies the stiffest direction under uniaxial tensile stress does not coincide with the stiffest direction under uniaxial tensile strain. Because uniaxial tensile stress tests are easier to conduct than uniaxial tensile strain tests, it is more convenient to obtain a standardized compliance matrix, and the discussions of this paper focus on compliance matrixes.

### *2.3. The number of independent elastic material constants for symmetric and two-dimensional materials*

When a material has some orientational symmetry, more material intrinsic directions can be used to set up a standardized coordinate system without the arbitrariness of the coordinate orientation. As demonstrated above, only under a standardized coordinate system, the number of independent elastic material constants can be correctly exhibited.

For a material with a single symmetry plane (or monoclinic symmetry), if $x_1$-axis is taken as the normal direction of the symmetric plane, and $x_2$ -axis is taken the stiffest direction in the plane, a standardized coordinate system is then set up. The corresponding standardized compliance matrix is

$$\mathbf{S}^{sym} = \begin{bmatrix} S_{11}^{sym} & S_{12}^{sym} & S_{13}^{sym} & S_{14}^{sym} & 0 & 0 \\ & S_{22}^{sym} & S_{23}^{sym} & 0 & 0 & 0 \\ & & S_{33}^{sym} & S_{34}^{sym} & 0 & 0 \\ & (sym) & & S_{44}^{sym} & 0 & 0 \\ & & & & S_{55}^{sym} & S_{56}^{sym} \\ & & & & & S_{66}^{sym} \end{bmatrix} \tag{9}$$

The superscript “sym” implies that the material orientational symmetry is used, and this compliance matrix is called as the **symmetry orientation based standardized compliance matrix,** or **SYM-OS compliance matrix** for abbreviation. In the above equation $S_{24}^{sym} = 0$ has been used according to Eq.(7), and noting that $S_{22}^{sym}$ is the minimum of $S_{22}$ in the symmetry plane. It is noted from Eq.(9) that a material with a single symmetry plane has only 12 independent elastic material constants, different from the statement in the textbooks that there are 13 independent elastic constants, since the arbitrariness of the coordinate orientation is usually not emphasized. Cowin(Cowin, 1995) also proved the appropriate selection of the coordinate system may reduce the number of the distinct constants for monoclinic symmetry by one.

For an orthotropic material with three orthogonal symmetry planes, $x_1$-axis and $x_2$ -axis are usually taken as two normal directions of the planes, and the three arbitrariness of the coordinate orientation are naturally excluded. The corresponding SYM-OS compliance matrix Eq.(2) can correctly exhibit 9 independent elastic material constants, as presented in the textbooks.

When only the two-dimensional anisotropic elasticity is under consideration, the sub-compliance-matrix $\mathbf{S}^{2D}$ has six components, i.e., $S_{11}, S_{22}, S_{33}, S_{12}, S_{16}$, and $S_{26}$. The corresponding STF-OS compliance matrix is easily obtained as

$$\hat{\mathbf{S}}^{2D} = \begin{bmatrix} \hat{S}_{11} & \hat{S}_{12} & 0 \\ \hat{S}_{12} & \hat{S}_{22} & \hat{S}_{26} \\ 0 & \hat{S}_{26} & \hat{S}_{66} \end{bmatrix} \tag{10}$$

This standardized compliance matrix clearly reveals that there are only 5 independent elastic material constants, not six.

*2.4. Comparison between two types of standardized compliance matrixes: the stiffest orientation based one and symmetry orientation based one*

As mentioned above, two types of material intrinsic directions, i.e. the direction with an extreme stiffness and the one related to some symmetries, can be used to establish a special coordinate system. The corresponding compliance matrix without the arbitrariness of coordinate system orientation is called a standardized one in this paper. If only the stiffest directions are adopted in establishing the special coordinate system, an STF-OS compliance matrix is obtained, while if any direction related to some symmetry is adopted, an SYM-OS compliance matrix is obtained.

The STF-OS compliance matrix and SYM-OS compliance matrix have different advantages and applications. The former exists for any anisotropic material and has a unique representation, which is convenient in comparing among different materials. The latter only exists for materials with orientational symmetry, but can exhibit the number of independent material constants more clearly than the former.

For materials with some symmetry planes, one might speculate that the special coordinate system based on the extreme stiffness direction coincides with the one based on the normal directions of the symmetry planes. This is not always true, and will be demonstrated by the following example. An overall orthotropic laminate material is composed of unidirectional fiber-reinforced laminas, as shown in Fig.4. The composite laminate has four laminas with equal thickness and the set of fiber orientations [40°, -40°, -40°, 40°].

The compliance matrix of the unidirectional fiber-reinforced lamina under its local principle coordinate system (aligned with the fiber direction) is supposed as

$$\mathbf{S}^{lamina} = \frac{1}{E_1^{lamina}} \begin{bmatrix} 1 & -0.3 & 0 \\ -0.3 & 100 & 0 \\ 0 & 0 & 2 \end{bmatrix} \tag{11}$$

According to the classical laminate theory, the in-plane strains for different laminas are the same, thus the compliance matrix for the laminate can be obtained. The distribution of values of $1/S_{11}$ for composite laminate in different directions is shown in Fig.4. It is found that the stiffest direction (or the direction with the maximum $1/S_{11}$) is not along any symmetric axis, $x_1-$axis or $x_2-$axis. Therefore, the STF-OS compliance matrix and the SYM-OS compliance matrix of this composite laminate are different.

## 3. Determining material orientational symmetry through the searching process of the stiffest orientation based standardized compliance matrix

Materials sometimes do not show their orientational symmetry in their appearance. In this case, we may firstly treat it as a general anisotropic material, and measure its compliance components in an arbitrarily chosen coordinate system. We then search for the STF-OS coordinate system and the corresponding STF-OS compliance matrix. During this searching process, some information, such as a sphere of orientation colored by the values of uniaxial tensile stiffness, can be obtained and may provide hints on the material orientational symmetry, but it is interesting to find that not all these hints can definitely lead to positive results, as discussed later in this section.

### *3.1. Symmetry investigation on materials with two stiffest orientation based standardized coordinate systems*

According to Condition I-IV in Section 2, any material has a unique STF-OS compliance matrix. For general anisotropic materials, this standardized compliance matrix can only be obtained under a single standardized coordinate system. If the STF-OS compliance matrix can be obtained under two STF-OS coordinate systems, we will prove that the material has a symmetry plane.

The two STF-OS coordinate systems are assumed $\hat{x}_1^{(1)}-\hat{x}_2^{(1)}-\hat{x}_3^{(1)}$ and $\hat{x}_1^{(2)}-\hat{x}_2^{(2)}-\hat{x}_3^{(2)}$, under which the STF-OS compliance matrixes $\hat{\mathbf{S}}$ are identical. For convenience, a reference coordinate system $\mathrm{X}_1-\mathrm{X}_2-\mathrm{X}_3$, as shown in Fig.5, is introduced. In order to show schematics clearly, the origins of different coordinate systems are sometimes translationally shifted in this paper. Without loss of generality, it is assumed that $\mathrm{X}_1$-axis is the bisector of angle $2\varphi$ between $\hat{x}_1^{(1)}$-axis and $\hat{x}_1^{(2)}$-axis, and $\mathrm{X}_2$-axis is also within the plane of $\hat{x}_1^{(1)}$-axis and $\hat{x}_1^{(2)}$-axis. The angle around $\hat{x}_1^{(1)}$-axis from $\mathrm{X}_1-\mathrm{X}_2$ plane to $\hat{x}_2^{(1)}$-axis and is denoted as $\Theta^{(1)}$, which takes a positive value when following the right hand rule. The angle around $\hat{x}_1^{(2)}$-axis from $\mathrm{X}_1-\mathrm{X}_2$ plane to $\hat{x}_2^{(2)}$-axis is denoted as $\Theta^{(2)}$.

Because $\hat{x}_1^{(1)}-\hat{x}_2^{(1)}-\hat{x}_3^{(1)}$ and $\hat{x}_1^{(2)}-\hat{x}_2^{(2)}-\hat{x}_3^{(2)}$ are only two STF-OS coordinate systems, it is easy to know that the rotations from the coordinate system $\hat{x}_1^{(1)}-\hat{x}_2^{(1)}-\hat{x}_3^{(1)}$ to $\hat{x}_1^{(2)}-\hat{x}_2^{(2)}-\hat{x}_3^{(2)}$ should be the same as those from the coordinate system $\hat{x}_1^{(2)}-\hat{x}_2^{(2)}-\hat{x}_3^{(2)}$ to $\hat{x}_1^{(1)}-\hat{x}_2^{(1)}-\hat{x}_3^{(1)}$, as shown in Fig.6. Therefore $\Theta^{(1)}$ should be equal to $\Theta^{(2)}$ (denoted by $\Theta$ later), and this relation can also been proved analytically as presented in Appendix B.

Through the same rotations (rotating $\Theta$ around first axis and then rotating $-\varphi$ around the third axis), $\hat{x}_1^{(1)}-\hat{x}_2^{(1)}-\hat{x}_3^{(1)}$ and $\hat{x}_1^{(2)}-\hat{x}_2^{(2)}-\hat{x}_3^{(2)}$ become $\mathrm{X}_1\text{-}\mathrm{X}_2\text{-}\mathrm{X}_3$ and $\mathrm{X}_1\text{-}\mathrm{X}_2^{(-)}\text{-}\mathrm{X}_3^{(-)}$, respectively, as shown in Fig.7. Obviously, the compliance matrixes under $X_1$-$X_2$-$X_3$ and $X_1$-$X_2^{(-)}$-$X_3^{(-)}$ coordinate systems are the same. It also noted that $X_1$-$X_2$-$X_3$ can be obtained by rotating $X_1$-$X_2^{(-)}$-$X_3^{(-)}$ 180 degree around $X_1$ -axis, which leads 8 opposite values in the components for two corresponding compliance matrixes, similar to Eq.(3a) and Eq.(3b). From these two relations, the compliance matrix under $X_1$-$X_2$-$X_3$ should therefore have 8 zeroes in the upper triangle as,

$$\mathbf{S}_{X_1\text{-}X_2\text{-}X_3} = \begin{bmatrix} S_{11} & S_{12} & S_{13} & S_{14} & 0 & 0 \\ & S_{22} & S_{23} & S_{24} & 0 & 0 \\ & & S_{33} & S_{34} & 0 & 0 \\ & (sym) & & S_{44} & 0 & 0 \\ & & & & S_{55} & S_{56} \\ & & & & & S_{66} \end{bmatrix} \tag{12}$$

which implies that $X_2 - X_3$ plane is a symmetry plane for this material.

*3.2. Identification of the material symmetry through the orientational variation of* $S_{11}$

According to Condition I for an STF-OS coordinate system, determining $\hat{x}_1$-axis needs to search for the global minimum $S_{11}$ over all possible orientations. Therefore, a sphere of orientation colored by the values of $S_{11}$ as shown in Fig.8 can be naturally obtained during the searching process. By observing this sphere, we can find the symmetry of $S_{11}$ if there is any. Our question is that, does the symmetry of $S_{11}$ definitely imply the symmetry of material elastic properties? In specific, three types of symmetries, i.e. mirror symmetry, transversely isotropic symmetry and isotropic symmetry, are discussed as follows. Euler angles $\alpha$, $\beta$ and $\gamma$ shown in Fig. 8(a) are adopted to describe spatial orientations with respect to a reference coordinate system $X_1$-$X_2$-$X_3$.

*A mirror symmetry in the orientational variation of* $S_{11}$

If a sphere of orientation colored by $S_{11}$ possesses a mirror symmetry, without loss of generality, we assume that $X_2$-$X_3$ plane is the symmetry plane, as shown in Fig.8(b). Obviously, for any $\alpha$ and $\beta$, the direction expressed by Euler angles $(\alpha,\beta)$ and the direction $(\alpha,\pi-\beta)$, are symmetrical about the $X_2$-$X_3$ plane, and their corresponding $S_{11}$ are the same. The mirror symmetry of $S_{11}$ can be analytically expressed as

$$S_{11}\left(\alpha,\beta,S_{ij}^{(0)}\right)=S_{11}\left(\alpha,\pi-\beta,S_{ij}^{(0)}\right) \tag{13}$$

where $S_{ij}^{(0)}$ are the components of the compliance matrix under $X_1$-$X_2$-$X_3$ coordinate system. From the relation between the original compliance matrix and the rotated one (see Appendix A), the detailed expression of Eq.(13) can be obtained as

$$\begin{aligned}
0=&S_{11}\left(\alpha,\beta,S_{ij}^{(0)}\right)-S_{11}\left(\alpha,\pi-\beta,S_{ij}^{(0)}\right)\\
=&-4\sin(\beta)\cos(\beta)\Big\{\cos^2(\beta)\left[\cos(\alpha)S_{15}^{(0)}-\sin(\alpha)S_{16}^{(0)}\right]\\
&+\cos(\alpha)\sin^2(\beta)\left[\sin^2(\alpha)\left(S_{25}^{(0)}+S_{46}^{(0)}\right)+\cos^2(\alpha)S_{35}^{(0)}\right]\\
&-\sin(\alpha)\sin^2(\beta)\left[\sin^2(\alpha)S_{26}^{(0)}+\cos^2(\alpha)\left(S_{36}^{(0)}+S_{45}^{(0)}\right)\right]\Big\}
\end{aligned} \tag{14}$$

Because the above equation holds for any $\alpha$ and $\beta$, it can be inferred that

$$S_{15}^{(0)}=S_{16}^{(0)}=S_{26}^{(0)}=S_{35}^{(0)}=S_{25}^{(0)}+S_{46}^{(0)}=S_{36}^{(0)}+S_{45}^{(0)}=0 \tag{15}$$

which are not consistent with the conditions as shown in Eq.(12) for the mirror symmetry about $X_2$-$X_3$ plane of material elastic properties. This implies that a material with the mirror symmetry of the tensile stiffness does not definitely possess the same symmetry in its elastic properties.

*A transversely isotropic symmetry in the orientational variation of* $S_{11}$

A transversely isotropic symmetry is a much higher symmetry than a mirror symmetry disused above. If a sphere of orientation colored by $S_{11}$ possesses a transversely isotropic symmetry, without loss of generality, $X_2$-$X_3$ plane is assumed to be the transversely isotropic plane, as shown in Fig.8(c). For any $\alpha$ and $\beta$, the direction expressed by Euler angles $(\alpha,\beta)$ and the direction $(0,\beta)$ have the same $S_{11}$, i.e.,

$$S_{11}\left(\alpha,\beta,S_{ij}^{(0)}\right)=S_{11}\left(0,\beta,S_{ij}^{(0)}\right) \tag{16}$$

From the relation between the original compliance matrix and the transformed one (see Appendix A), the detailed expression of Eq.(16) can be obtained as Eq.(17).

$$\begin{aligned}
0=&S_{11}\left(\alpha,\beta,S_{ij}^{(0)}\right)-S_{11}\left(0,\beta,S_{ij}^{(0)}\right)\\
=&\sin^4(\alpha)\sin^4(\beta)S_{22}^{(0)}+\left(\cos^4(\alpha)-1\right)\sin^4(\beta)S_{33}^{(0)}+\sin^2(\alpha)\cos^2(\alpha)\sin^4(\beta)\left(S_{44}^{(0)}+2S_{23}^{(0)}\right)\\
&-\sin^2(\beta)\cos^2(\beta)\sin^2(\alpha)\left(S_{55}^{(0)}-S_{66}^{(0)}-2\left(S_{12}^{(0)}-S_{13}^{(0)}\right)\right)\\
&-2\sin^2(\beta)\cos^2(\beta)\sin(\alpha)\cos(\alpha)\left(S_{14}^{(0)}+S_{56}^{(0)}\right)-2\sin^3(\beta)\cos(\beta)\sin^2(\alpha)\cos(\alpha)\left(S_{25}^{(0)}+S_{46}^{(0)}\right)\\
&+2\sin^3(\beta)\cos(\beta)\sin(\alpha)\cos^2(\alpha)\left(S_{36}^{(0)}+S_{45}^{(0)}\right)+2\sin(\beta)\cos^3(\beta)\left(1-\cos(\alpha)\right)S_{15}^{(0)}\\
&+2\sin(\beta)\cos^3(\beta)\sin(\alpha)S_{16}^{(0)}-2\sin^4(\beta)\sin^3(\alpha)\cos(\alpha)S_{24}^{(0)}+2\sin^3(\beta)\cos(\beta)\sin^3(\alpha)S_{26}^{(0)}\\
&-2\sin^4(\beta)\sin(\alpha)\cos^3(\alpha)S_{34}^{(0)}+2\sin^3(\beta)\cos(\beta)\left(1-\cos^3(\alpha)\right)S_{35}^{(0)}
\end{aligned} \tag{17}$$

For any $\alpha$ and $\beta$, the following relation can satisfy Eq.(17),

$$\begin{aligned}
&S_{22}^{(0)}=S_{33}^{(0)},\ S_{55}^{(0)}-S_{66}^{(0)}=2\left(S_{12}^{(0)}-S_{13}^{(0)}\right),\ S_{44}^{(0)}=2\left(S_{22}^{(0)}-S_{23}^{(0)}\right)\\
&S_{14}^{(0)}+S_{56}^{(0)}=S_{25}^{(0)}+S_{46}^{(0)}=S_{36}^{(0)}+S_{45}^{(0)}=0\\
&S_{15}^{(0)}=S_{16}^{(0)}=S_{24}^{(0)}=S_{26}^{(0)}=S_{34}^{(0)}=S_{35}^{(0)}=0
\end{aligned} \tag{18}$$

However, Eq.(18) is not consistent with the following conditions for the transversely isotropic symmetry with the isotropic $X_2$ - $X_3$ plane on material elastic properties ,

$$
\begin{aligned}
& S_{22}^{(0)} = S_{33}^{(0)}, S_{55}^{(0)} = S_{66}^{(0)}, S_{12}^{(0)} = S_{13}^{(0)},\ S_{44}^{(0)} = 2\left(S_{22}^{(0)} - S_{23}^{(0)}\right) \\
& S_{14}^{(0)} = S_{56}^{(0)} = S_{25}^{(0)} = S_{46}^{(0)} = S_{36}^{(0)} = S_{45}^{(0)} = 0 \\
& S_{15}^{(0)} = S_{16}^{(0)} = S_{24}^{(0)} = S_{26}^{(0)} = S_{34}^{(0)} = S_{35}^{(0)} = 0
\end{aligned} \tag{19}
$$

This also implies that a material with the transversely isotropic symmetry of the tensile stiffness does not definitely possess the same symmetry in its elastic properties.

<u>*An isotropic symmetry in the orientational variation of* $S_{11}$</u>

An isotropic symmetry is the highest symmetry. If a sphere of orientation colored by $S_{11}$ possesses an isotropic symmetry, any orientational plane is a transversely isotropic plane, and it is easy to know that the symmetry with two orthogonal transversely isotropic planes is equivalent to an isotropic symmetry. Therefore, adopting $X_2$ - $X_3$ plane and $X_1$ - $X_2$ plane as the transversely isotropic planes, and using Eq.(18), the following relation can be derived

$$
\begin{aligned}
& S_{11}^{(0)} = S_{22}^{(0)} = S_{33}^{(0)}, \quad S_{15}^{(0)} = S_{16}^{(0)} = S_{24}^{(0)} = S_{26}^{(0)} = S_{34}^{(0)} = S_{35}^{(0)} = 0, \\
& S_{14}^{(0)} + S_{56}^{(0)} = S_{25}^{(0)} + S_{46}^{(0)} = S_{36}^{(0)} + S_{45}^{(0)} = 0, \\
& S_{44}^{(0)} = 2\left(S_{13}^{(0)} - S_{23}^{(0)}\right), \ S_{55}^{(0)} = 2\left(S_{11}^{(0)} - S_{13}^{(0)}\right), \ S_{66}^{(0)} = 2\left(S_{11}^{(0)} - S_{12}^{(0)}\right),
\end{aligned} \tag{20}
$$

or expressed in the compliance matrix form

$$
\mathbf{S}_{X_1\text{-}X_2\text{-}X_3} = \begin{bmatrix}
S_{11}^{(0)} & S_{12}^{(0)} & S_{13}^{(0)} & S_{14}^{(0)} & 0 & 0 \\
 & S_{11}^{(0)} & S_{23}^{(0)} & 0 & S_{25}^{(0)} & 0 \\
 & & S_{11}^{(0)} & 0 & 0 & S_{36}^{(0)} \\
 & \text{(sym)} & & 2\left(S_{11}^{(0)} - S_{23}^{(0)}\right) & -S_{36}^{(0)} & -S_{25}^{(0)} \\
 & & & & 2\left(S_{11}^{(0)} - S_{13}^{(0)}\right) & -S_{14}^{(0)} \\
 & & & & & 2\left(S_{11}^{(0)} - S_{12}^{(0)}\right)
\end{bmatrix} \tag{21}
$$

Obviously, Eq. (21) is not consistent with the following compliance matrix for the isotropic symmetry of material elastic properties.

$$
\mathbf{S}^{isotropic} = \begin{bmatrix}
S_{11}^{(0)} & S_{12}^{(0)} & S_{12}^{(0)} & 0 & 0 & 0 \\
 & S_{11}^{(0)} & S_{12}^{(0)} & 0 & 0 & 0 \\
 & & S_{11}^{(0)} & 0 & 0 & 0 \\
 & \text{(sym)} & & 2\left(S_{11}^{(0)} - S_{12}^{(0)}\right) & 0 & 0 \\
 & & & & 2\left(S_{11}^{(0)} - S_{12}^{(0)}\right) & 0 \\
 & & & & & 2\left(S_{11}^{(0)} - S_{12}^{(0)}\right)
\end{bmatrix} \tag{22}
$$

It is very interesting to note from Eq. (21) and Eq. (22) that a material with isotropic tensile stiffness is not definitely an isotropic elastic material. In another word, the symmetry of the material elastic properties cannot be completely identified by only measuring the tensile stiffness of all directions. More interestingly, these statements are only correct for three dimensional cases, but do not hold for two dimensional anisotropy/materials. Namely, an isotropic symmetry of tensile stiffness in two dimensions is definitely equivalent to an isotropic symmetry of material elastic properties. The detailed analytic proof on two dimensional cases is presented in Appendix C.

For three-dimensional materials with an isotropic symmetry of tensile stiffness, Figure 9 shows a schematic example, a cuboid RVE subject to a uniaxial tensile stress $\sigma_{11}$. From the interpretation context related to Fig.3, we know $S_{15} = S_{16} = 0$, since every direction is the stiffest direction when the RVE possesses isotropic symmetry of tensile stiffness. However, a non-vanishing shear deformation $\gamma_{23}$ normal to the tensile stress $\sigma_{11}$ cannot be excluded by those conditions, resulting in a non-vanishing compliance component $S_{14}$ and anisotropy of the material.

## 4. A simple and correct measure of anisotropy degree

The last part of the previous section exhibits a possibility that, at least in mathematical sense, a material with isotropic tensile stiffness does not have isotropic elastic properties. However, many researchers have incorrectly adopted the ratio between the maximum and minimum tensile stiffnesses as the measure of anisotropy degree (e.g., Ni and Chiang, 2007; Lirkov et al., 2014). In the following, we will construct a periodical lattice structure to demonstrate the failure of this measure in identifying the anisotropy.

### *4.1. An anisotropic material with isotropic tensile stiffness*

Figure 10 shows a periodic cubic unit cell (or a representative volume element) of a lattice structure with side length *l*, which consists of two-force bars and connecting nodes. One node is inside the cube and eight other nodes are located at the corners. An optimization is carried out on the stiffnesses of bars and the position of the inside node to make its tensile stiffness isotropic while keeping the anisotropy of material elasticity, and the results are given in Table 1. The corresponding effective compliance matrix $\mathbf{S}$ of this latticed material is given in $X_1 - X_2 - X_3$ coordinate system as

$$\mathbf{S} = \frac{l}{k^{ab}} \begin{bmatrix} 0.1736 & -0.0434 & -0.0426 & -0.0111 & 0 & 0 \\ & 0.1736 & -0.0436 & 0 & -0.0066 & 0 \\ & & 0.1736 & 0 & 0 & 0.0080 \\ & (sym) & & 0.4399 & -0.0080 & 0.0066 \\ & & & & 0.4378 & 0.0111 \\ & & & & & 0.4395 \end{bmatrix} \quad (23)$$

where $k^{ab}$ is the stiffness of the bar connecting node a and node b.

Obviously, the tensile stiffnesses along $X_1$, $X_2$ and $X_3$-direction are the same, but this compliance matrix is not an isotropic one. Furthermore, the finite element software ABAQUS is adopted to test the isotropy of the tensile stiffness along any direction. Four arbitrary directions denoted by Euler angles ( $\alpha,\beta,\gamma$ as shown in Fig.8) are chosen to simulate the uniaxial tensile tests, and the stresses and strains in both the reference $X_1 - X_2 - X_3$ coordinate system and corresponding rotated coordinate systems are given in Table 2. It is easy to apply loading in the reference $X_1 - X_2 - X_3$ coordinate system and to obtain the tensile stiffness in the rotated coordinate system. The first component of the strain matrix in the rotated coordinate system clearly exhibits the isotropy of the tensile stiffness. Hence, this is a counter example that demonstrates the invalidity of the measure of anisotropy degree based on the ratio between the maximum and minimum tensile stiffnesses. A more proper measure is needed.

*4.2. A measure of anisotropy degree based on the maximum shear-extension coupling coefficient*

As mentioned in Section 3, a non-vanishing shear deformation $\gamma_{23}$ cannot be excluded in a material with only isotropic tensile stiffness. In this paper, we therefore propose the maximum ratio between the shear and tensile strain under uniaxial tension in all orientations, i.e. the maximum shear-extension coupling coefficient (Jones, 1998; Lekhnitskii et al., 1964), as a simple measure of anisotropy degree,

$$\eta_{\max} = \max\left(\frac{\gamma_{normal}}{\varepsilon_{tensile}}, \frac{\gamma_{in-plane}}{\varepsilon_{tensile}}\right)_{for\ all\ orientations} . \tag{24}$$

where $\gamma_{normal} = \gamma_{23}$ and $\gamma_{in-plane} = \gamma_{12}$ are shear strains normal to and in the plane of the uniaxial tensile direction ( $X_1$ - direction), respectively, and $\varepsilon_{tensile} = \varepsilon_{11}$ is the tensile strain. $\eta_{\max}$ can also be expressed with components of compliance matrix as

$$\eta_{\max} = \max\left(\frac{S_{14}}{S_{11}}, \frac{S_{15}}{S_{11}}\right)_{for\ all\ coordinate\ systems} \tag{25}$$

When a material subject to uniaxial stretcthing in any direction has no shear deformation, $\eta_{\max}$ is zero, and we will prove in the following that the material has isotropic elaticity.

According to the relevant analysis of Fig.3 and Eq.(5)(6), no in-plane shear strain for uniaxial stretcthing in any direction implies that the derivative of the tensile stiffness along any direction with respect to the orientation angle is zero, and the material has isotropic tensile stiffness. Therefore, the corresponding compliance matrix has the form as Eq.(21). Still based on zero shear strain,

$$S_{14}^{(0)} = S_{25}^{(0)} = S_{36}^{(0)} = 0 \tag{26}$$

The compliance matrix becomes

$$
\mathbf{S}_{X_1\text{-}X_2\text{-}X_3} = \begin{bmatrix} S_{11}^{(0)} & S_{12}^{(0)} & S_{13}^{(0)} & 0 & 0 & 0 \\ & S_{11}^{(0)} & S_{23}^{(0)} & 0 & 0 & 0 \\ & & S_{11}^{(0)} & 0 & 0 & 0 \\ & \text{(sym)} & & 2\left(S_{11}^{(0)} - S_{23}^{(0)}\right) & 0 & 0 \\ & & & & 2\left(S_{11}^{(0)} - S_{13}^{(0)}\right) & 0 \\ & & & & & 2\left(S_{11}^{(0)} - S_{12}^{(0)}\right) \end{bmatrix} \tag{27}
$$

If $S_{12}^{(0)} \neq S_{13}^{(0)}$, $\varepsilon_{22} \neq \varepsilon_{33}$ for the material under $X_1$-direction tension, and non-zero shear strain will appear when the coordinate system rotates around $X_1$-axis. Therefore, $S_{12}^{(0)} = S_{13}^{(0)}$. Similarly, $S_{12}^{(0)} = S_{23}^{(0)}$. Eq.(27) is then the same as the isotropic elastic compliance matrix Eq.(22). It is proved that this maximum shear-extension coupling coefficient $\eta_{\max}$ is a correct and simple measure of anisotropy degree.

## 5. Conclusion

In this paper, standardized compliance matrixes for general anisotropic materials and a new measure of anisotropy degree have been proposed and discussed. The following conclusions are achieved.

(1) For a general anisotropic material, a unique stiffest orientation based standardized (STF-OS) compliance matrix is proposed, which clearly exhibits only 18 (not 21) independent elastic material constants and makes it convenient to compare the elastic properties between different materials.

(2) If a material has some orientational symmetry, the symmetry orientation based standardized (SYM-OS) compliance matrix can also be obtained, which can show the reduced independent elastic material constants. During the searching process for the symmetry, it is interesting to find that a material with some symmetry of the tensile stiffness, such as isotropic symmetry, does not definitely possess the same symmetry in its elastic properties.

(3) It is important to find that the ratio between the maximum and minimum tensile stiffnesses is not a correct measure of anisotropy degree. The maximum shear-extension coupling coefficient in all orientations is proposed to be a simple and correct measure of anisotropy degree. Therefore, no shear-extension coupling is the most essential feature of isotropic elasticity, not isotropic tensile stiffness.

**Acknowledgement**

The authors acknowledge the support from National Natural Science Foundation of China (Grant Nos. 11425208, 51232004 and 11372158) and Tsinghua University Initiative Scientific Research Program (No. 2011Z02173).The authors are grateful for helpful discussions with Professor QuanShui Zheng.

**References**

Boulanger, P., Hayes, M., 1995. On Young's modulus for anisotropic media. Journal of applied mechanics 62, 819-820.

Bower, A.F., 2011. Applied mechanics of solids. CRC press.
Cowin, S., Mehrabadi, M., 1995. Anisotropic symmetries of linear elasticity. Applied Mechanics Reviews 48, 247-285.
Cowin, S.C., 1995. On the number of distinct elastic constants associated with certain anisotropic elastic symmetries, Theoretical, Experimental, and Numerical Contributions to the Mechanics of Fluids and Solids. Springer, pp. 210-224.
Fedorov, F.I., 2012. Theory of Elastic Waves in Crystals. Springer US.
Herakovich, C.T., 2012. Mechanics of composites: a historical review. Mechanics Research Communications 41, 1-20.
Jones, R.M., 1998. Mechanics of composite materials. CRC Press.
Kaw, A.K., 2005. Mechanics of composite materials. CRC press.
Lekhnitskii, S., Fern, P., Brandstatter, J.J., Dill, E., 1964. Theory of elasticity of an anisotropic elastic body. Physics Today 17, 84.
Ting, T.C., Horgan, C., 1996. Anisotropic elasticity: theory and applications. Journal of Applied Mechanics 63, 1056.
Truesdell, C., 1966. Existence of longitudinal waves. The Journal of the Acoustical Society of America 40, 729-730.
Truesdell, C., 1968. Comment on Longitudinal Waves. The Journal of the Acoustical Society of America 43, 170-170.

## Appendix A Rotation relation from a reference coordinate system to a rotated one

In Euclidean space $E^3$ (with Cartesian metric), when rotating the reference coordinate system $x_1^{(0)}-x_2^{(0)}-x_3^{(0)}$ to $x_1^{(1)}-x_2^{(1)}-x_3^{(1)}$, the basis of the vector space $\left\{\mathbf{e}_1^{(0)},\mathbf{e}_2^{(0)},\mathbf{e}_3^{(0)}\right\}$ becomes $\left\{\mathbf{e}_1^{(1)},\mathbf{e}_2^{(1)},\mathbf{e}_3^{(1)}\right\}$. The orthogonal matrix is denoted as $\mathbf{\Omega}$ and can be written as

$$\mathbf{\Omega}=\begin{bmatrix} \mathbf{e}_1^{(1)}\cdot\mathbf{e}_1^{(0)} & \mathbf{e}_1^{(1)}\cdot\mathbf{e}_2^{(0)} & \mathbf{e}_1^{(1)}\cdot\mathbf{e}_3^{(0)} \\ \mathbf{e}_2^{(1)}\cdot\mathbf{e}_1^{(0)} & \mathbf{e}_2^{(1)}\cdot\mathbf{e}_2^{(0)} & \mathbf{e}_2^{(1)}\cdot\mathbf{e}_3^{(0)} \\ \mathbf{e}_3^{(1)}\cdot\mathbf{e}_1^{(0)} & \mathbf{e}_3^{(1)}\cdot\mathbf{e}_2^{(0)} & \mathbf{e}_3^{(1)}\cdot\mathbf{e}_3^{(0)} \end{bmatrix} \tag{A.1}$$

Considering the rotations of coordinate system $x_1^{(0)}-x_2^{(0)}-x_3^{(0)}$ around $x_1,x_2,x_3$-axis through an arbitrary angle $\theta_1,\theta_2,\theta_3$, respectively, the corresponding transformation matrixes are

$$\boldsymbol{\Omega}(\theta_1)=\begin{bmatrix}1 & 0 & 0\\ 0 & \cos(\theta_1) & \sin(\theta_1)\\ 0 & -\sin(\theta_1) & \cos(\theta_1)\end{bmatrix},\qquad \boldsymbol{\Omega}(\theta_2)=\begin{bmatrix}\cos(\theta_2) & 0 & -\sin(\theta_2)\\ 0 & 1 & 0\\ \sin(\theta_2) & 0 & \cos(\theta_2)\end{bmatrix},$$
$$\boldsymbol{\Omega}(\theta_3)=\begin{bmatrix}\cos(\theta_3) & \sin(\theta_3) & 0\\ -\sin(\theta_3) & \cos(\theta_3) & 0\\ 0 & 0 & 1\end{bmatrix}. \tag{A.2}$$

The compliance matrix $\boldsymbol{\sigma}^{(0)}$ in coordinate system $x_1^{(0)}-x_2^{(0)}-x_3^{(0)}$ is

$$\boldsymbol{\sigma}^{(0)}=\begin{bmatrix}\sigma_{11}^{(0)} & \sigma_{12}^{(0)} & \sigma_{13}^{(0)}\\ \sigma_{12}^{(0)} & \sigma_{22}^{(0)} & \sigma_{23}^{(0)}\\ \sigma_{13}^{(0)} & \sigma_{23}^{(0)} & \sigma_{33}^{(0)}\end{bmatrix} \tag{A.3}$$

And the compliance matrix $\boldsymbol{\sigma}^{(1)}$ in coordinate system $x_1^{(1)}-x_2^{(1)}-x_3^{(1)}$ can be computed by $\boldsymbol{\sigma}^{(0)}$ as

$$\boldsymbol{\sigma}^{(1)}=\boldsymbol{\Omega}\cdot\boldsymbol{\sigma}^{(0)}\cdot\boldsymbol{\Omega}^T \tag{A.4}$$

When rotating the coordinate system $x_1^{(0)}-x_2^{(0)}-x_3^{(0)}$ around $x_1$-axis through an angle $\theta_1$, the components of $\boldsymbol{\sigma}^{(1)}$ can be obtained through Eq.(A.2)(A.3)(A.4).

$$\begin{aligned}
\sigma_{11}^{(1)} &= \sigma_{11}^{(0)}\\
\sigma_{22}^{(1)} &= \cos^2(\theta_1)\sigma_{22}^{(0)}+\sin^2(\theta_1)\sigma_{33}^{(0)}+2\sin(\theta_1)\cos(\theta_1)\sigma_{23}^{(0)}\\
\sigma_{33}^{(1)} &= \sin^2(\theta_1)\sigma_{22}^{(0)}+\cos^2(\theta_1)\sigma_{33}^{(0)}-2\sin(\theta_1)\cos(\theta_1)\sigma_{23}^{(0)}\\
\sigma_{23}^{(1)} &= -\sin(\theta_1)\cos(\theta_1)\sigma_{22}^{(0)}+\sin(\theta_1)\cos(\theta_1)\sigma_{33}^{(0)}+\left(\cos^2(\theta_1)-\sin^2(\theta_1)\right)\sigma_{23}^{(0)}\\
\sigma_{13}^{(1)} &= \cos(\theta_1)\sigma_{13}^{(0)}-\sin(\theta_1)\sigma_{12}^{(0)}\\
\sigma_{12}^{(1)} &= \sin(\theta_1)\sigma_{13}^{(0)}+\cos(\theta_1)\sigma_{12}^{(0)}
\end{aligned} \tag{A.5}$$

The components of $\boldsymbol{\sigma}^{(1)}$ in Eq.(A.5) can be also expressed as

$$\sigma_{ij}^{(1)}=\left(T_{\theta_1}^{\sigma}\right)_{ik}\sigma_{kj}^{(0)}, \tag{A.6}$$

where $\mathbf{T}_{\theta_1}^{\boldsymbol{\sigma}}$ is the stress transformation matrix when rotating the coordinate system $x_1^{(0)}-x_2^{(0)}-x_3^{(0)}$ around $x_1$-axis through an angle $\theta_1$ and can be written as

$$\mathbf{T}_{\theta_1}^{\boldsymbol{\sigma}} = \begin{bmatrix} 1 & 0 & 0 & 0 & 0 & 0 \\ 0 & \cos^2(\theta_1) & \sin^2(\theta_1) & 2\sin(\theta_1)\cos(\theta_1) & 0 & 0 \\ 0 & \sin^2(\theta_1) & \cos^2(\theta_1) & -2\sin(\theta_1)\cos(\theta_1) & 0 & 0 \\ 0 & -\sin(\theta_1)\cos(\theta_1) & \sin(\theta_1)\cos(\theta_1) & \cos^2(\theta_1)-\sin^2(\theta_1) & 0 & 0 \\ 0 & 0 & 0 & 0 & \cos(\theta_1) & -\sin(\theta_1) \\ 0 & 0 & 0 & 0 & \sin(\theta_1) & \cos(\theta_1) \end{bmatrix}. \quad (A.7)$$

Similarly, the corresponding strain transformation matrix $\mathbf{T}_{\theta_1}^{\boldsymbol{\varepsilon}}$ can be obtained as

$$\mathbf{T}_{\theta_1}^{\boldsymbol{\varepsilon}} = \begin{bmatrix} 1 & 0 & 0 & 0 & 0 & 0 \\ 0 & \cos^2(\theta_1) & \sin^2(\theta_1) & \sin(\theta_1)\cos(\theta_1) & 0 & 0 \\ 0 & \sin^2(\theta_1) & \cos^2(\theta_1) & -\sin(\theta_1)\cos(\theta_1) & 0 & 0 \\ 0 & -2\sin(\theta_1)\cos(\theta_1) & 2\sin(\theta_1)\cos(\theta_1) & \cos^2(\theta_1)-\sin^2(\theta_1) & 0 & 0 \\ 0 & 0 & 0 & 0 & \cos(\theta_1) & -\sin(\theta_1) \\ 0 & 0 & 0 & 0 & \sin(\theta_1) & \cos(\theta_1) \end{bmatrix}. \quad (A.8)$$

When rotating the coordinate system $x_1^{(0)} - x_2^{(0)} - x_3^{(0)}$ around $x_2$-axis through an angle $\theta_2$, the stress and strain transformation matrixes, $\mathbf{T}_{\theta_2}^{\boldsymbol{\sigma}}$ and $\mathbf{T}_{\theta_2}^{\boldsymbol{\varepsilon}}$, can be written as

$$\mathbf{T}_{\theta_2}^{\boldsymbol{\sigma}} = \begin{bmatrix} \cos^2(\theta_2) & 0 & \sin^2(\theta_2) & 0 & -2\sin(\theta_2)\cos(\theta_2) & 0 \\ 0 & 1 & 0 & 0 & 0 & 0 \\ \sin^2(\theta_2) & 0 & \cos^2(\theta_2) & 0 & 2\sin(\theta_2)\cos(\theta_2) & 0 \\ 0 & 0 & 0 & \cos(\theta_2) & 0 & \sin(\theta_2) \\ \sin(\theta_2)\cos(\theta_2) & 0 & -\sin(\theta_2)\cos(\theta_2) & 0 & \cos^2(\theta_2)-\sin^2(\theta_2) & 0 \\ 0 & 0 & 0 & -\sin(\theta_2) & 0 & \cos(\theta_2) \end{bmatrix}$$

$$\mathbf{T}_{\theta_2}^{\boldsymbol{\varepsilon}} = \begin{bmatrix} \cos^2(\theta_2) & 0 & \sin^2(\theta_2) & 0 & -\sin(\theta_2)\cos(\theta_2) & 0 \\ 0 & 1 & 0 & 0 & 0 & 0 \\ \sin^2(\theta_2) & 0 & \cos^2(\theta_2) & 0 & \sin(\theta_2)\cos(\theta_2) & 0 \\ 0 & 0 & 0 & \cos(\theta_2) & 0 & \sin(\theta_2) \\ 2\sin(\theta_2)\cos(\theta_2) & 0 & -2\sin(\theta_2)\cos(\theta_2) & 0 & \cos^2(\theta_2)-\sin^2(\theta_2) & 0 \\ 0 & 0 & 0 & -\sin(\theta_2) & 0 & \cos(\theta_2) \end{bmatrix} \quad (A.9)$$

When rotating the coordinate system $x_1^{(0)} - x_2^{(0)} - x_3^{(0)}$ around $x_3$-axis through an angle $\theta_3$, the stress and strain transformation matrixes, $\mathbf{T}_{\theta_3}^{\boldsymbol{\sigma}}$ and $\mathbf{T}_{\theta_3}^{\boldsymbol{\varepsilon}}$, can be written as

$$\mathbf{T}_{\theta_3}^{\boldsymbol{\sigma}} = \begin{bmatrix} \cos^2(\theta_3) & \sin^2(\theta_3) & 0 & 0 & 0 & 2\sin(\theta_3)\cos(\theta_3) \\ \sin^2(\theta_3) & \cos^2(\theta_3) & 0 & 0 & 0 & -2\sin(\theta_3)\cos(\theta_3) \\ 0 & 0 & 1 & 0 & 0 & 0 \\ 0 & 0 & 0 & \cos(\theta_3) & -\sin(\theta_3) & 0 \\ 0 & 0 & 0 & \sin(\theta_3) & \cos(\theta_3) & 0 \\ -\sin(\theta_3)\cos(\theta_3) & \sin(\theta_3)\cos(\theta_3) & 0 & 0 & 0 & \cos^2(\theta_3)-\sin^2(\theta_3) \end{bmatrix}$$

$$\mathbf{T}_{\theta_3}^{\boldsymbol{\varepsilon}} = \begin{bmatrix} \cos^2(\theta_3) & \sin^2(\theta_3) & 0 & 0 & 0 & \sin(\theta_3)\cos(\theta_3) \\ \sin^2(\theta_3) & \cos^2(\theta_3) & 0 & 0 & 0 & -\sin(\theta_3)\cos(\theta_3) \\ 0 & 0 & 1 & 0 & 0 & 0 \\ 0 & 0 & 0 & \cos(\theta_3) & -\sin(\theta_3) & 0 \\ 0 & 0 & 0 & \sin(\theta_3) & \cos(\theta_3) & 0 \\ -2\sin(\theta_3)\cos(\theta_3) & 2\sin(\theta_3)\cos(\theta_3) & 0 & 0 & 0 & \cos^2(\theta_3)-\sin^2(\theta_3) \end{bmatrix} \quad \text{(A.10)}$$

When rotating the coordinate system $x_1^{(0)}-x_2^{(0)}-x_3^{(0)}$ around $x_i$-axis through an angle $\theta_i$ ($i=1,2,3$), the stress matrix $\boldsymbol{\sigma}^{(1)}$ and strain matrix $\boldsymbol{\varepsilon}^{(1)}$ can be obtained according to Eq.(A.7)-(A.10).

$$\boldsymbol{\sigma}^{(1)} = \mathbf{T}_{\theta_i}^{\boldsymbol{\sigma}} \cdot \boldsymbol{\sigma}^{(0)}, \qquad \boldsymbol{\varepsilon}^{(1)} = \mathbf{T}_{\theta_i}^{\boldsymbol{\varepsilon}} \cdot \boldsymbol{\varepsilon}^{(0)} \quad \text{(A.11)}$$

Since the generalized Hooke's law,

$$\boldsymbol{\varepsilon}^{(0)} = \mathbf{S}^{(0)} \cdot \boldsymbol{\sigma}^{(0)}, \qquad \boldsymbol{\varepsilon}^{(1)} = \mathbf{S}^{(1)} \cdot \boldsymbol{\sigma}^{(1)} \quad \text{(A.12)}$$

the relation between compliance matrixes $\mathbf{S}^{(0)}$ and $\mathbf{S}^{(1)}$ can be obtained through Eq. (A.11)(A.12).

$$\mathbf{S}^{(1)} = \mathbf{T}_{\theta_i}^{\boldsymbol{\varepsilon}} \cdot \mathbf{S}^{(0)} \cdot \left(\mathbf{T}_{\theta_i}^{\boldsymbol{\sigma}}\right)^{-1} \quad \text{(A.13)}$$

As discussed in Section 2.1, Coordinate system (b), (c) and (d) can be obtained through rotating Coordinate system (a) $\pi$ angle around $\hat{x}_1^a$, $\hat{x}_2^a$ and $\hat{x}_3^a$-axis, respectively. Therefore, the compliance matrixes $\mathbf{S}^b, \mathbf{S}^c$ and $\mathbf{S}^d$ can be expressed as

$$\begin{aligned} \mathbf{S}^b &= \mathbf{T}_{\theta_1=\pi}^{\boldsymbol{\varepsilon}} \cdot \mathbf{S}^a \cdot \left(\mathbf{T}_{\theta_1=\pi}^{\boldsymbol{\sigma}}\right)^{-1} \\ \mathbf{S}^c &= \mathbf{T}_{\theta_2=\pi}^{\boldsymbol{\varepsilon}} \cdot \mathbf{S}^a \cdot \left(\mathbf{T}_{\theta_2=\pi}^{\boldsymbol{\sigma}}\right)^{-1} \\ \mathbf{S}^d &= \mathbf{T}_{\theta_3=\pi}^{\boldsymbol{\varepsilon}} \cdot \mathbf{S}^a \cdot \left(\mathbf{T}_{\theta_3=\pi}^{\boldsymbol{\sigma}}\right)^{-1} \end{aligned} \quad \text{(A.14)}$$

## Appendix B The demonstration of $\Theta^{(1)} = \Theta^{(2)}$

From Fig.6, the process of rotating the coordinate system $\hat{x}_1^{(1)}-\hat{x}_2^{(1)}-\hat{x}_3^{(1)}$ to $\hat{x}_1^{(2)} - \hat{x}_2^{(2)} - \hat{x}_3^{(2)}$ is as follows: firstly, rotating coordinate system $\hat{x}_1^{(1)} - \hat{x}_2^{(1)} - \hat{x}_3^{(1)}$ around $\hat{x}_1^{(1)}$ -axis through an angle $-\Theta^{(1)}$ ; then, rotating coordinate system $\hat{x}_1^{(1)} - x_2^{(1)} - x_3^{(1)}$ around $x_3^{(1)}$-axis through an angle $-2\varphi$;finally, rotating coordinate system $\hat{x}_1^{(2)} - x_2^{(1)'} - x_3^{(1)}$ around $\hat{x}_1^{(2)}$-axis through an angle $\pi + \Theta^{(2)}$ . After these rotations, according to the transformation relation of compliance matrixes under different coordinate systems shown in Eq.(A.13), the identical compliance matrix $\hat{\mathbf{S}}$ under the two STF-OS coordinate systems must satisfy the followings:

$$\hat{\mathbf{S}} = \left(\mathbf{T}^{\boldsymbol{\varepsilon}}_{\theta_1=\pi+\Theta^{(2)}}\right)\cdot\left(\mathbf{T}^{\boldsymbol{\varepsilon}}_{\theta_3=-2\varphi}\right)\cdot\left(\mathbf{T}^{\boldsymbol{\varepsilon}}_{\theta_1=-\Theta^{(1)}}\right)\cdot\hat{\mathbf{S}}\cdot\left(\mathbf{T}^{\boldsymbol{\sigma}}_{\theta_1=-\Theta^{(1)}}\right)^{-1}\cdot\left(\mathbf{T}^{\boldsymbol{\sigma}}_{\theta_3=-2\varphi}\right)^{-1}\cdot\left(\mathbf{T}^{\boldsymbol{\sigma}}_{\theta_1=\pi+\Theta^{(2)}}\right)^{-1} \quad \text{(B.1)}$$

Similarly, when rotating the coordinate system $\hat{x}_1^{(2)} - \hat{x}_2^{(2)} - \hat{x}_3^{(2)}$ to $\hat{x}_1^{(1)} - \hat{x}_2^{(1)} - \hat{x}_3^{(1)}$, the identical compliance matrix $\hat{\mathbf{S}}$ under the two STF-OS coordinate systems must satisfy the followings:

$$\hat{\mathbf{S}} = \left(\mathbf{T}^{\boldsymbol{\varepsilon}}_{\theta_1=\pi+\Theta^{(1)}}\right)\cdot\left(\mathbf{T}^{\boldsymbol{\varepsilon}}_{\theta_3=-2\varphi}\right)\cdot\left(\mathbf{T}^{\boldsymbol{\varepsilon}}_{\theta_1=-\Theta^{(2)}}\right)\cdot\hat{\mathbf{S}}\cdot\left(\mathbf{T}^{\boldsymbol{\sigma}}_{\theta_1=-\Theta^{(2)}}\right)^{-1}\cdot\left(\mathbf{T}^{\boldsymbol{\sigma}}_{\theta_3=-2\varphi}\right)^{-1}\cdot\left(\mathbf{T}^{\boldsymbol{\sigma}}_{\theta_1=\pi+\Theta^{(1)}}\right)^{-1} \quad \text{(B.2)}$$

According to Eq.(B.1)(B.2), it can be referred that

$$\Theta^{(1)}=\Theta^{(2)} \quad \text{(B.3)}$$

## Appendix C Identification of the material isotropic symmetry through the orientational variation of $S_{11}$ for two-dimensional materials

Supposed that a two-dimensional material is in the plane $X_1 - X_2$ and $X_3$-axis is perpendicular to the plane $X_1 - X_2$, the orientational variation of $S_{11}$ for two-dimensional materials in the plane $X_1 - X_2$ is isotropic. When rotating the coordinate system $X_1 - X_2$ around $X_3$-axis through angle $\gamma$, according to Eq. (A.13), the rotated compliance matrix $\mathbf{S}'^{2D}$ can be written as

$$\mathbf{S}'^{2D} = \left(\mathbf{T}^{\boldsymbol{\varepsilon}}_{\theta_3=\gamma}\right)\cdot\mathbf{S}^{2D}\cdot\left(\mathbf{T}^{\boldsymbol{\sigma}}_{\theta_3=\gamma}\right)^{-1} \quad \text{(C.1)}$$

When rotating the reference coordinate system $X_1 - X_2$ around $X_3$-axis through any angle $\gamma$, the value of component $S_{11}$ is always the same, i.e.

$$S'^{2D}_{11}\left(\theta_3=\gamma\right)=S^{2D}_{11} \tag{C.2}$$

It can be expressed in detail as

$$\begin{aligned}0=S'^{2D}_{11}-S^{2D}_{11}&=\left(\cos^4(\gamma)-1\right)S^{2D}_{11}+\sin^4(\gamma)S^{2D}_{22}+2\sin(\gamma)\cos^3(\gamma)S^{2D}_{16}\\&+2\sin^3(\gamma)\cos(\gamma)S^{2D}_{26}+\sin^2(\gamma)\cos^2(\gamma)\left(S^{2D}_{66}+2S^{2D}_{12}\right)\end{aligned} \tag{C.3}$$

When $\gamma=\pi/2$, it can be easily referred that $S^{2D}_{11}=S^{2D}_{22}$. Thus, Equation (C.3) can be simplified as

$$0=\sin(\gamma)\cos(\gamma)\left(S^{2D}_{66}+2S^{2D}_{12}-S^{2D}_{11}\right)+2\cos^2(\gamma)S^{2D}_{16}+2\sin^2(\gamma)S^{2D}_{26} \tag{C.4}$$

Then, the following can be referred

$$S^{2D}_{11}=S^{2D}_{22},\quad S^{2D}_{66}=2\left(S^{2D}_{11}-S^{2D}_{22}\right),\quad S^{2D}_{16}=S^{2D}_{26}=0 \tag{C.5}$$

This implies that a two-dimensional material with the isotropic symmetry of the tensile stiffness definitely possess the same isotropic symmetry in its elastic properties.

**Figure captions**

Figure 1 The four coordinate systems satisfying Condition I-III.
Figure 2 The specified sequence of compliance matrix components.
Figure 3 The deformation of RVE subject to a uniaxial tensile stress $\sigma_{11}$ along the $\hat{x}_1$-direction: (a) parallelogram deformation;(b) rectangle deformation.

Figure 4 The distribution of values of $1/S_{11}$ for composite laminate in different directions.

Figure 5 The two STF-OS coordinate systems with identical STF-OS compliance matrixes.

Figure 6 The rotation relation between coordinate systems $\hat{x}_1^{(1)} - \hat{x}_2^{(1)} - \hat{x}_3^{(1)}$ and $\hat{x}_1^{(2)} - \hat{x}_2^{(2)} - \hat{x}_3^{(2)}$.

Figure 7 The coordinate systems through the same rotations from coordinate systems $\hat{x}_1^{(1)} - \hat{x}_2^{(1)} - \hat{x}_3^{(1)}$ and $\hat{x}_1^{(2)} - \hat{x}_2^{(2)} - \hat{x}_3^{(2)}$.

Figure 8 The symmetry properties in the orientational variation of $S_{11}$ denoted as Euler angles: (a) Euler angles $\alpha$, $\beta$ and $\gamma$; (b) mirror symmetry; (c) transversely isotropic symmetry; (d) isotropic symmetry.

Figure 9 The deformation of a cuboid RVE subject to a uniaxial tensile stress $\sigma_{11}$.

Figure 10 A periodic cubic unit cell of a lattice structure with two-force bars and connecting nodes.

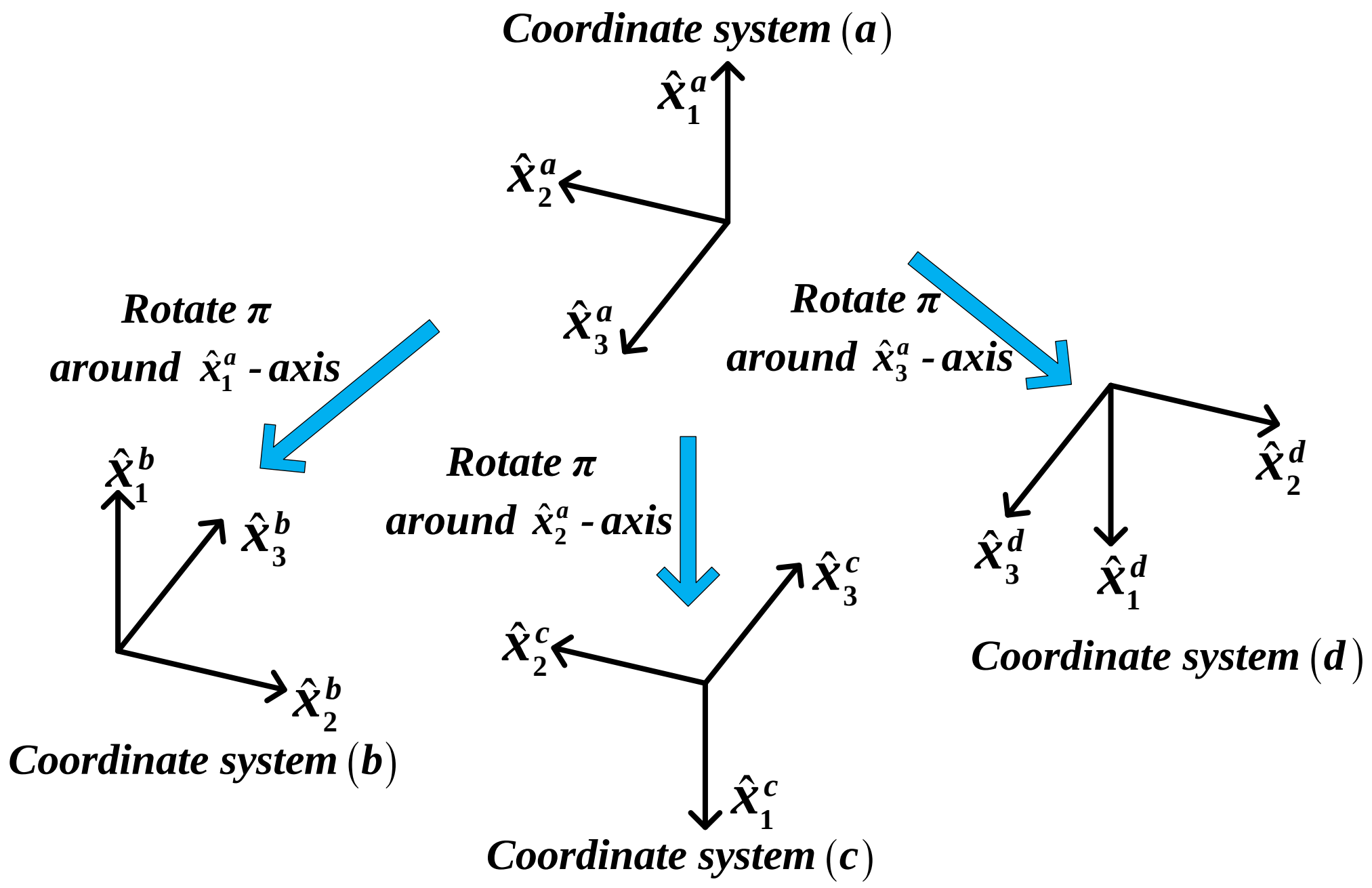

Fig. 1 The four coordinate systems satisfying Condition I-III.

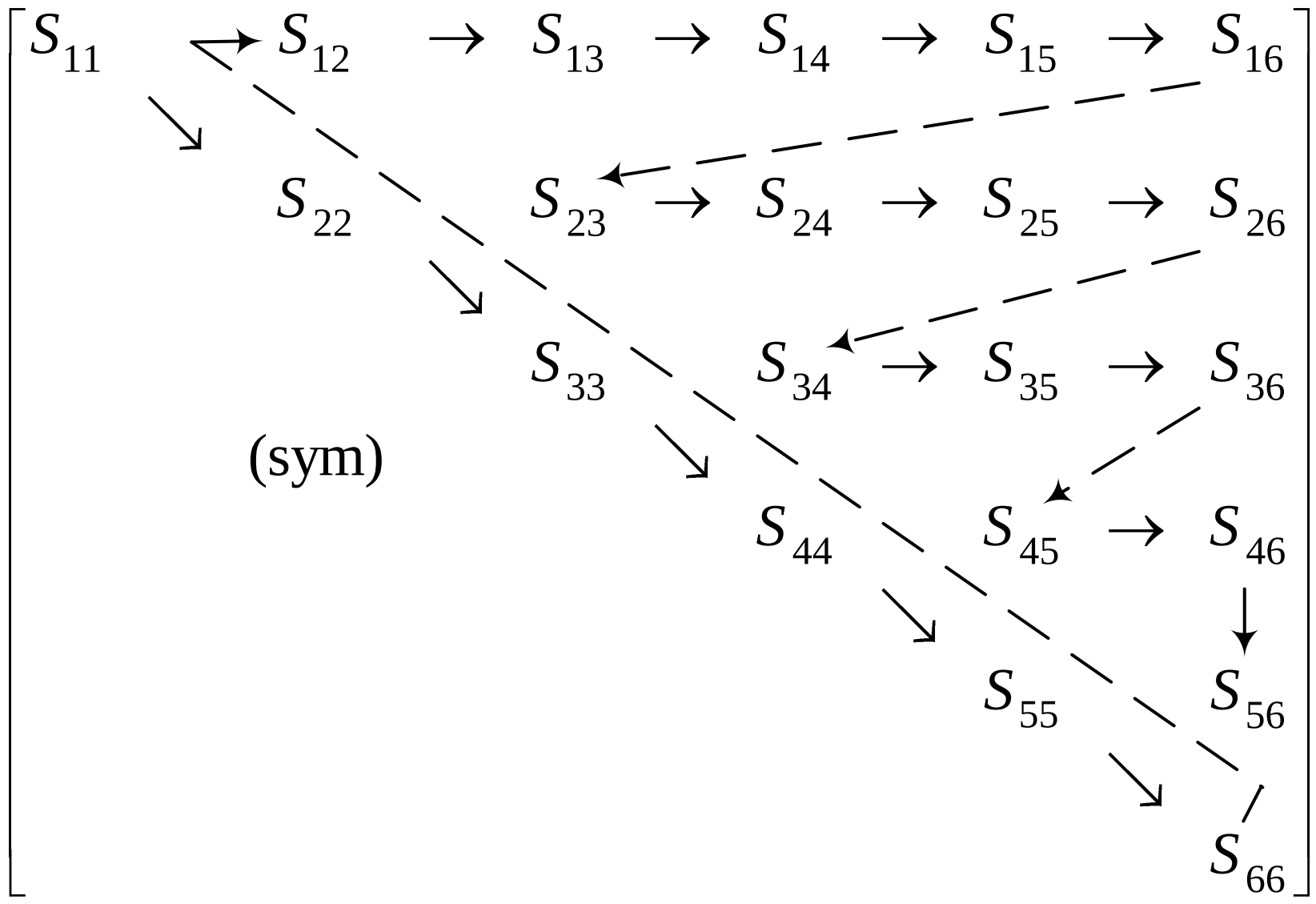

Fig. 2 The specified sequence of compliance matrix components.

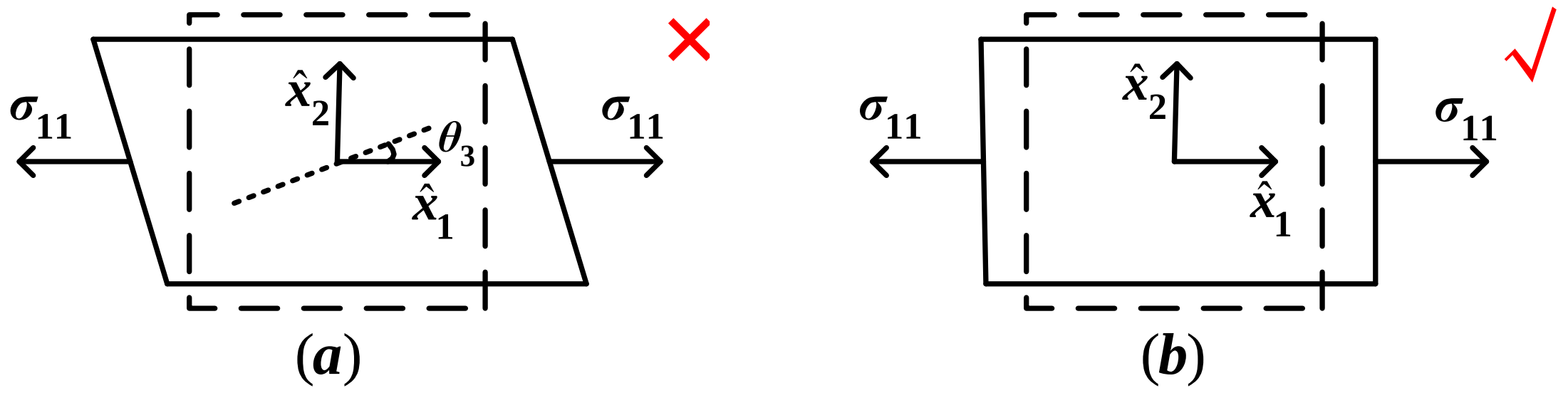

Fig. 3 The deformation of RVE subject to a uniaxial tensile stress $\sigma_{11}$ along the $\hat{x}_1$-direction: (a) parallelogram deformation;(b) rectangle deformation.

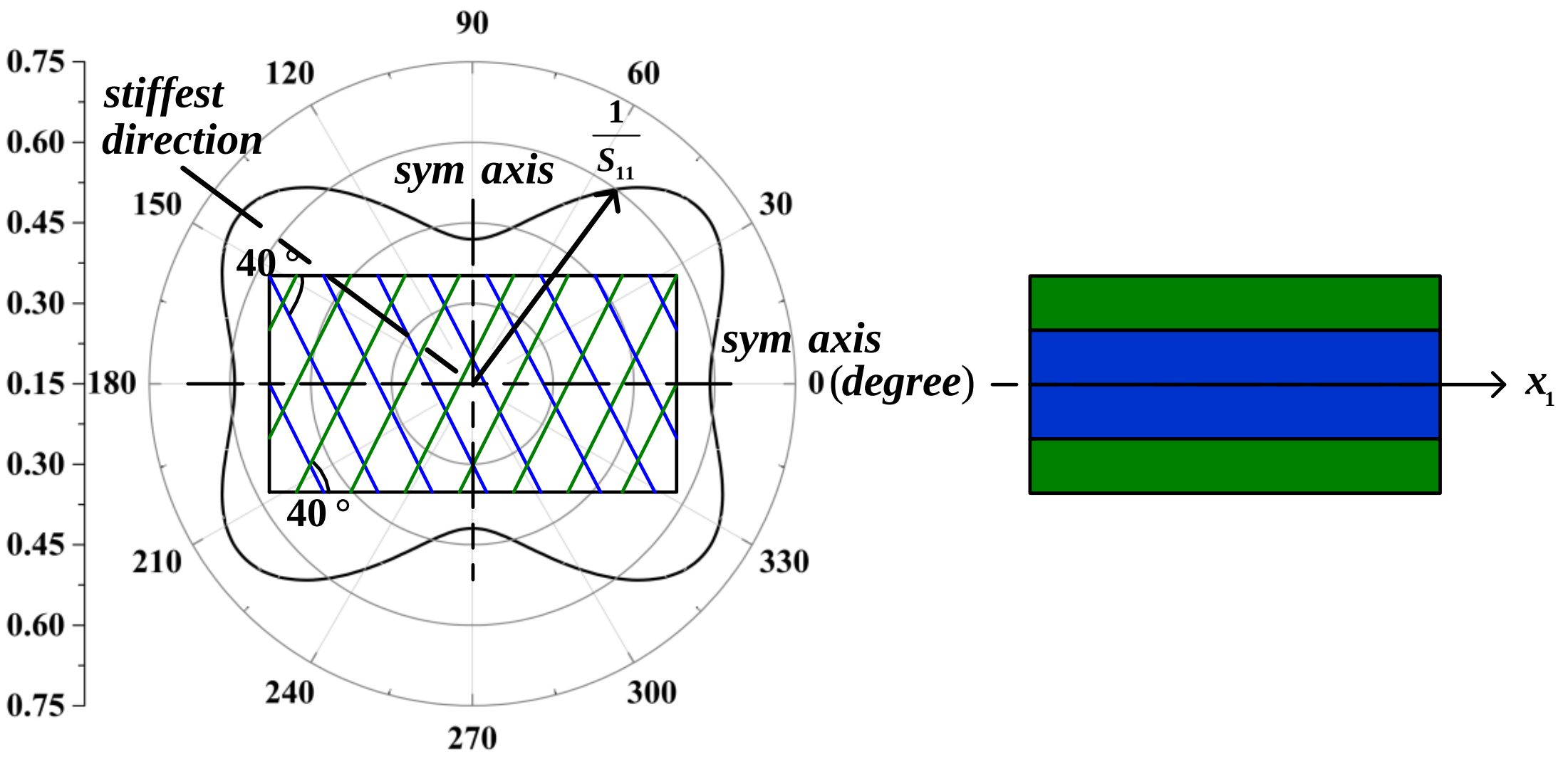

Fig. 4 The distribution of values of $1/S_{11}$

for composite laminate in different directions.

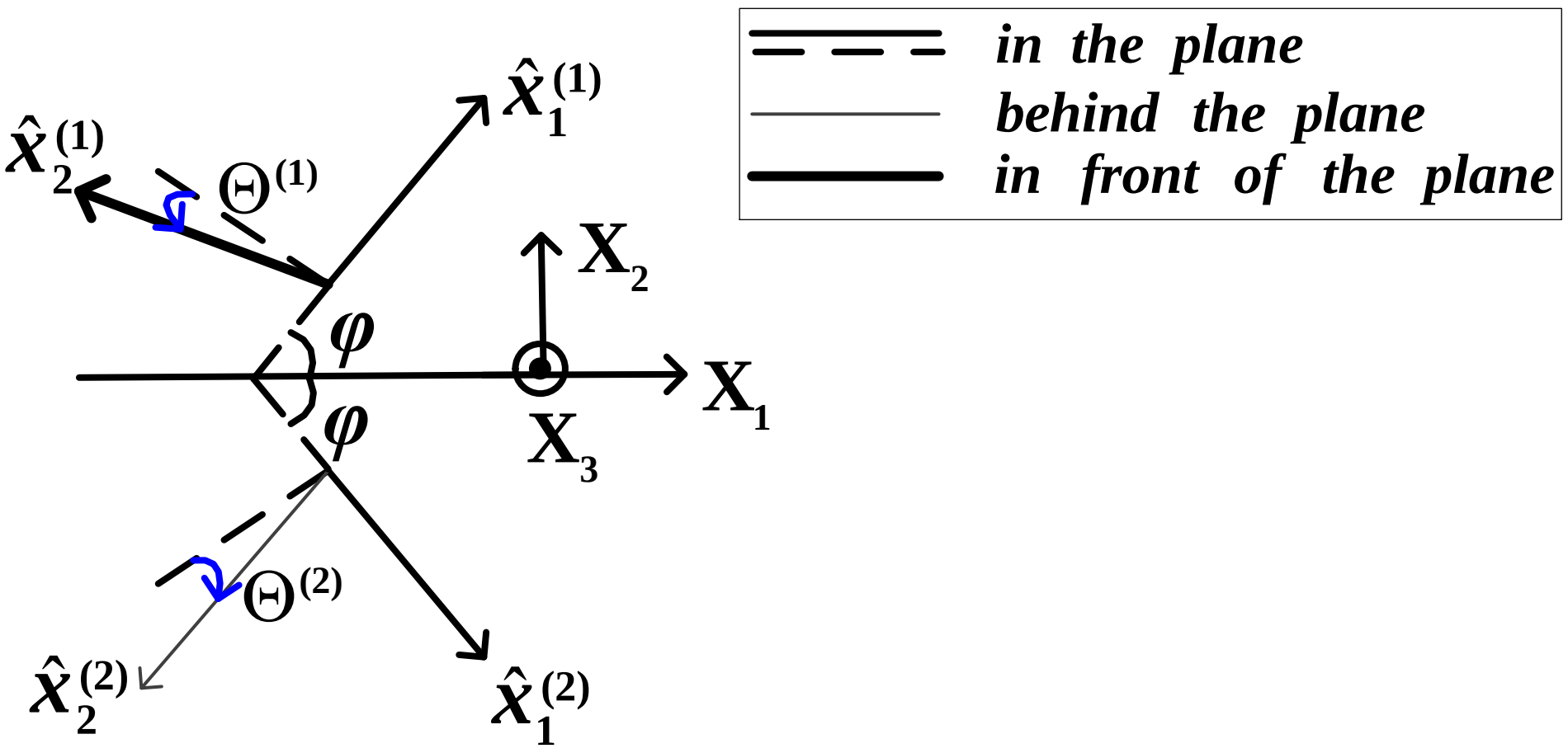

Fig. 5 The two STF-OS coordinate systems with identical STF-OS compliance matrixes.

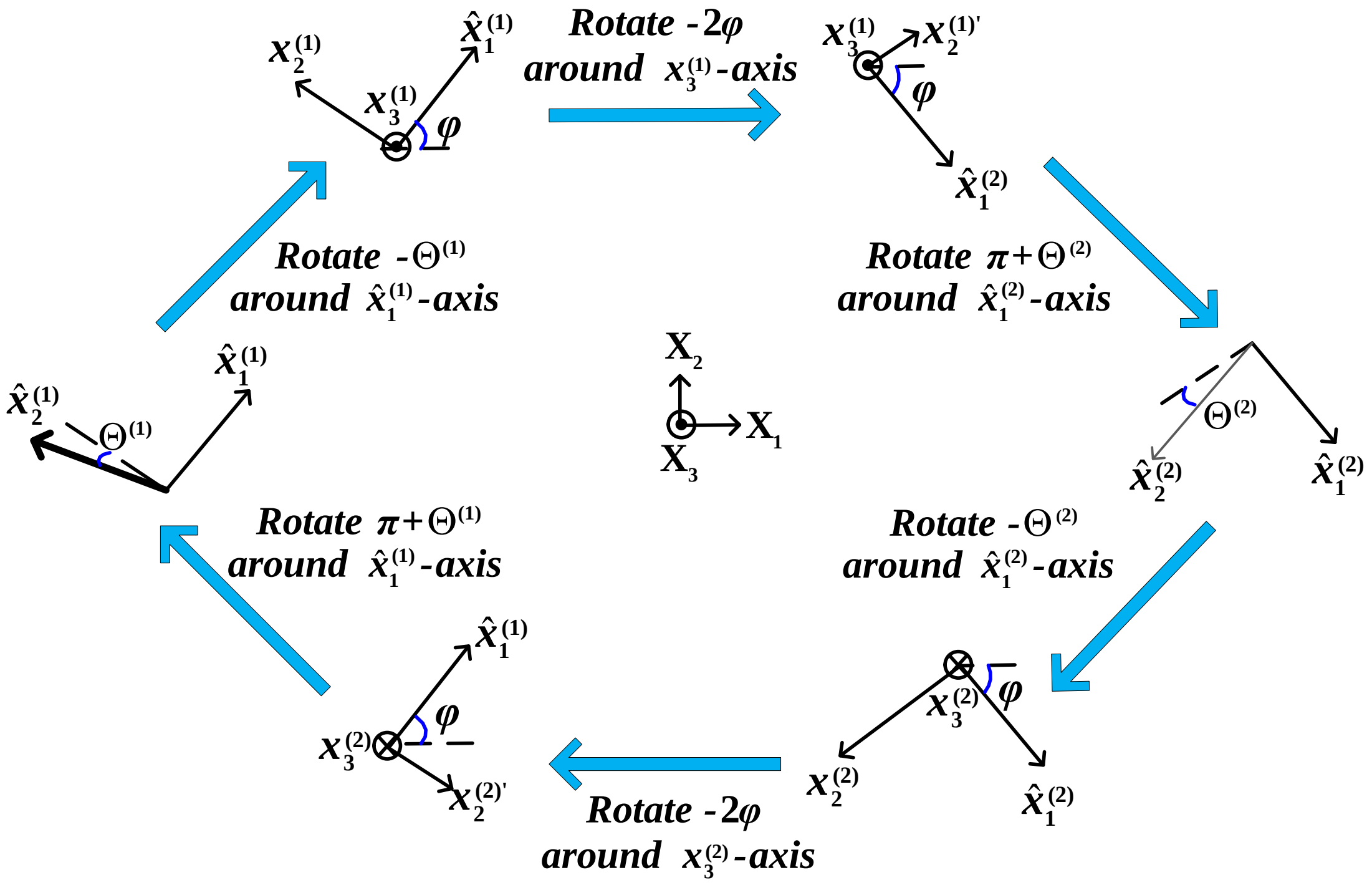

Fig. 6 The rotation relation between coordinate systems
$\hat{x}_1^{(1)}-\hat{x}_2^{(1)}-\hat{x}_3^{(1)}$ and $\hat{x}_1^{(2)}-\hat{x}_2^{(2)}-\hat{x}_3^{(2)}$.

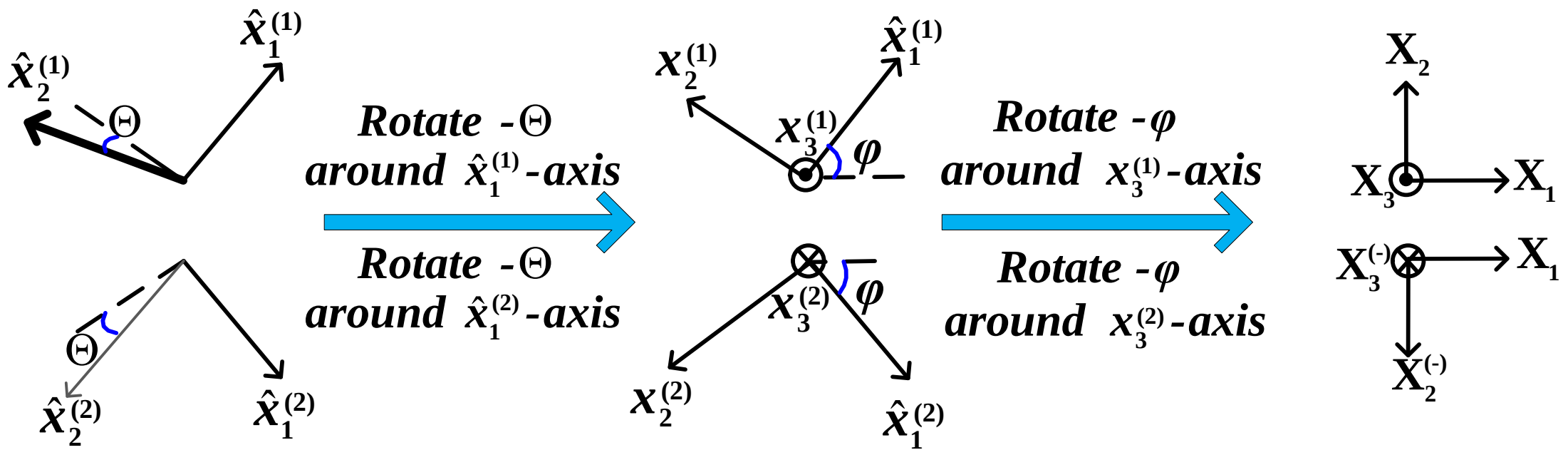

Fig. 7 The coordinate systems through the same rotations from coordinate systems $\hat{x}_1^{(1)}-\hat{x}_2^{(1)}-\hat{x}_3^{(1)}$ and $\hat{x}_1^{(2)}-\hat{x}_2^{(2)}-\hat{x}_3^{(2)}$.

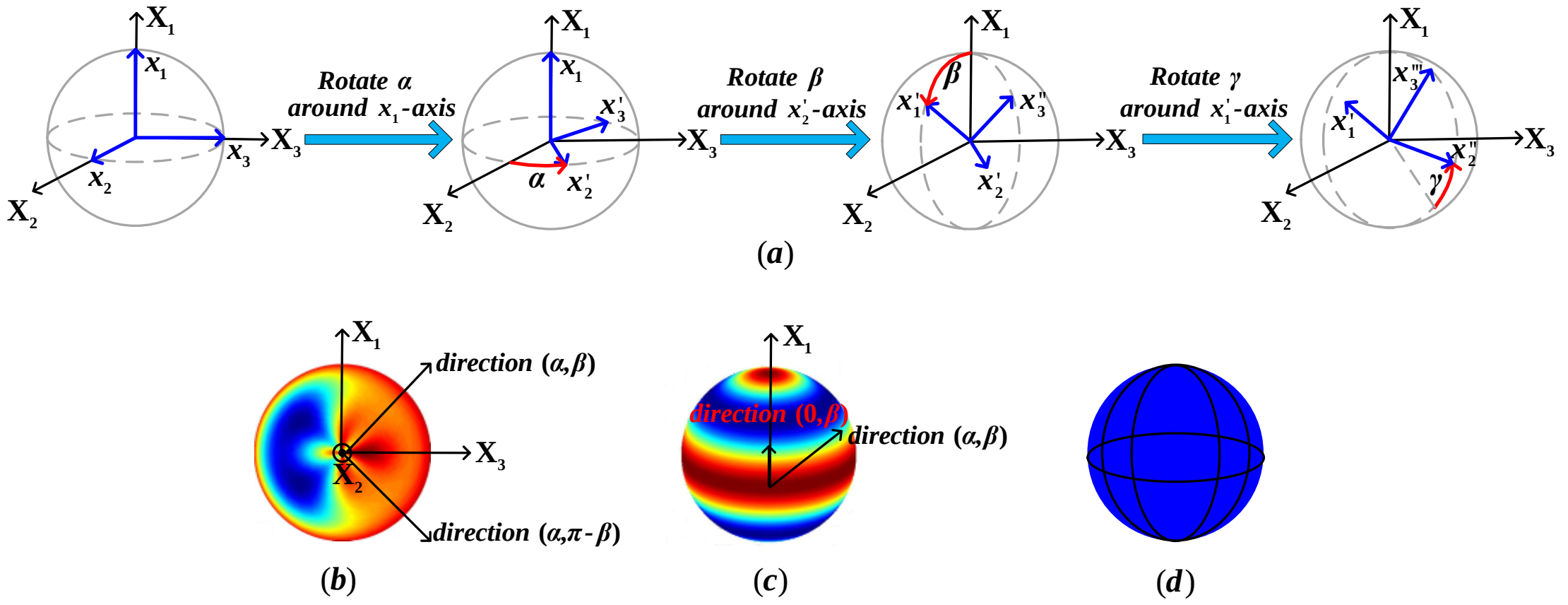

Fig. 8 The symmetry properties in the orientational variation of $S_{11}$ denoted as Euler angles: (a) Euler angles $\alpha$, $\beta$ and $\gamma$; (b) mirror symmetry; (c) transversely isotropic symmetry; (d) isotropic symmetry.

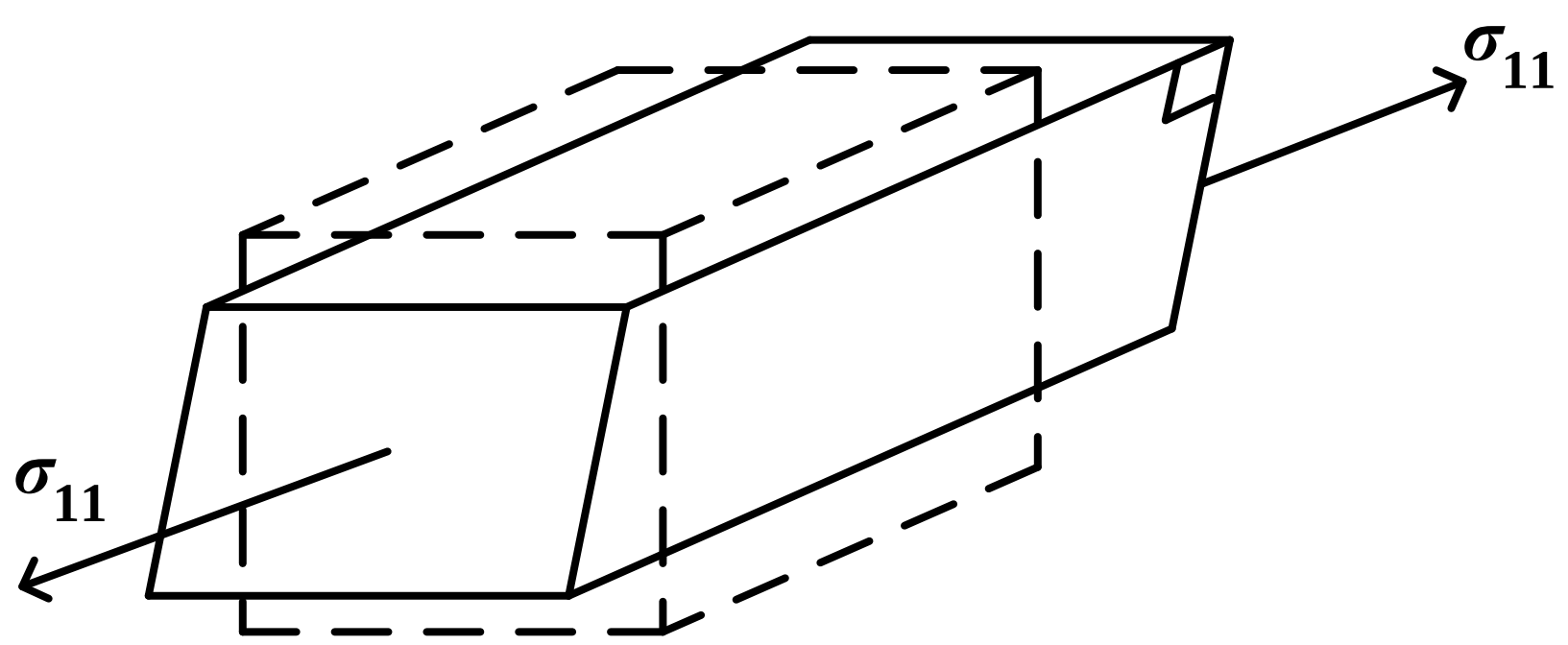

Fig.9 The deformation of a cuboid RVE subject to a uniaxial tensile stress $\sigma_{11}$ .

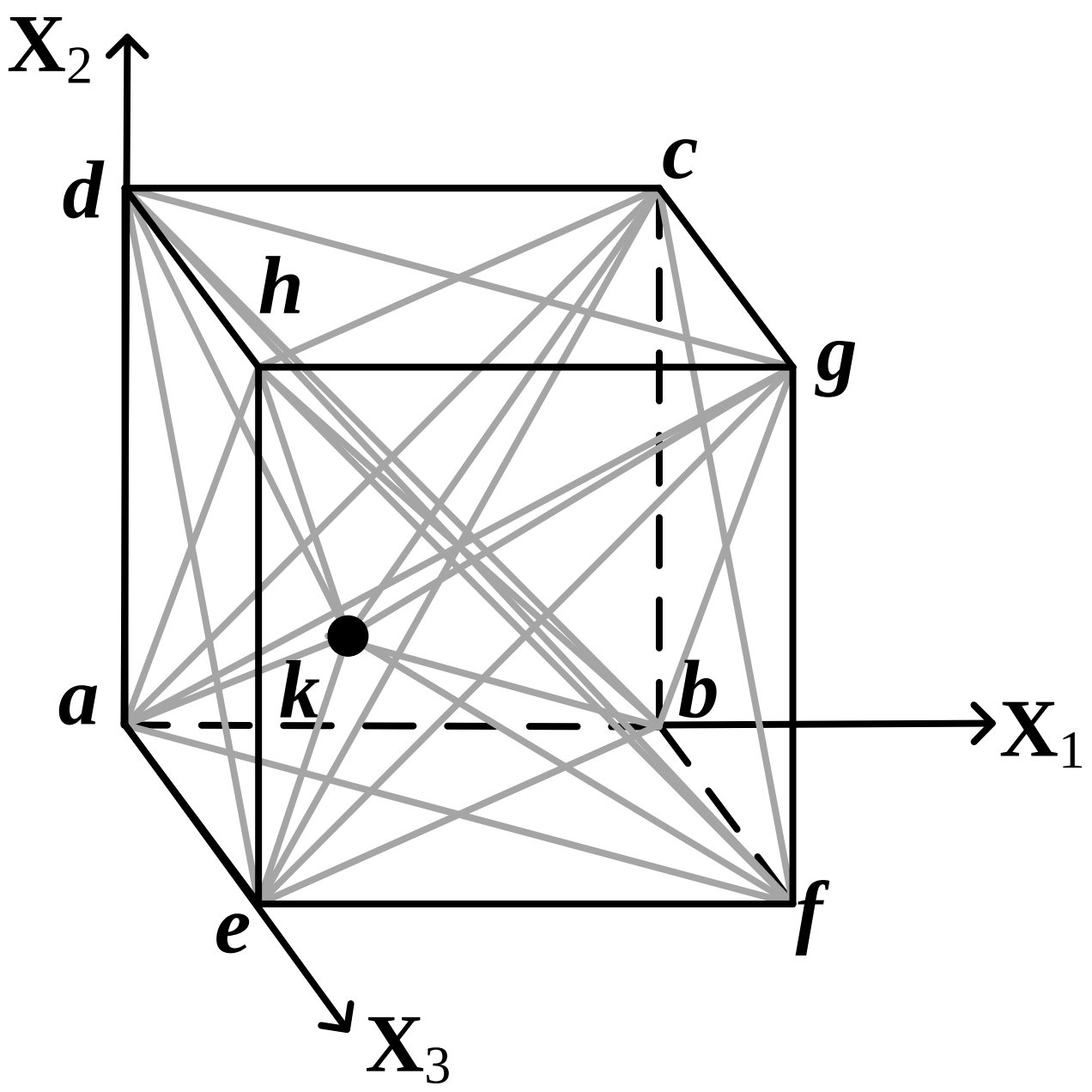

Fig. 10 A periodic cubic unit cell of a lattice structure with two-force bars and connecting nodes.

Table 1 The optimization results of the stiffnesses of bars and the position of the inside node

| | | | |
|---|---|---|---|
| The position of the inside node | $\left(\frac{x_1^k}{l}, \frac{x_2^k}{l}, \frac{x_3^k}{l}\right)$=(0.7866, 0.6736, 0.2795) | | |
| The bar stiffness | $k^{ab} = k^{ab}$ | $k^{ac} = 0.3727k^{ab}$ | $k^{ak} = 0.6465k^{ab}$ |
| | $k^{bc} = 0.9566k^{ab}$ | $k^{bd} = 0.2918k^{ab}$ | $k^{bk} = 1.8602k^{ab}$ |
| | $k^{ae} = 0.9781k^{ab}$ | $k^{ch} = 0.3782k^{ab}$ | $k^{ck} = 1.1756k^{ab}$ |
| | $k^{ag} = 0.8179k^{ab}$ | $k^{dg} = 0.3114k^{ab}$ | $k^{dk} = 1.6885k^{ab}$ |
| | $k^{bh} = 0.8451k^{ab}$ | $k^{ah} = 0.2173k^{ab}$ | $k^{ek} = 1.6332k^{ab}$ |
| | $k^{ce} = 0.4116k^{ab}$ | $k^{de} = 0.3331k^{ab}$ | $k^{fk} = 1.0263k^{ab}$ |
| | $k^{df} = 0.7512k^{ab}$ | | $k^{gk} = 2.0571k^{ab}$ |
| | | | $k^{hk} = 0.5734k^{ab}$ |

Table 2 The stresses and strains in $X_1 - X_2 - X_3$ coordinate system and corresponding rotated coordinate systems under the uniaxial tensile tests

| The direction of applied uniaxial stress | The components in $X_1 - X_2 - X_3$ coordinate system | | The components in rotated coordinate system with Euler angles $\alpha,\beta,\gamma$ | |
|---|---|---|---|---|
| | strain | stress | strain | stress |
| $\alpha = 0.2$ $\beta = 1.1$ $\gamma = 0.0$ | $\boldsymbol{\varepsilon} = 1\% \begin{bmatrix} 0.1003 \\ -0.0357 \\ 0.0322 \\ -0.0306 \\ -0.3100 \\ 0.0578 \end{bmatrix}_{X_1-X_2-X_3}$ | $\boldsymbol{\sigma} = \frac{0.01k^{ab}}{l} \begin{bmatrix} 0.6496 \\ 0.0138 \\ 0.3366 \\ -0.0682 \\ -0.7128 \\ 0.1445 \end{bmatrix}_{X_1-X_2-X_3}$ | $\boldsymbol{\varepsilon} = 1\% \begin{bmatrix} 0.1763 \\ -0.0390 \\ -0.0404 \\ -0.0052 \\ 0 \\ 0 \end{bmatrix}_{\alpha=0.2, \beta=1.1, \gamma=0.0}$ | $\boldsymbol{\sigma} = \frac{0.01k^{ab}}{l} \begin{bmatrix} 1 \\ 0 \\ 0 \\ 0 \\ 0 \\ 0 \end{bmatrix}_{\alpha=0.2, \beta=1.1, \gamma=0.0}$ |
| $\alpha = 0.5$ $\beta = 0.7$ $\gamma = 0.0$ | $\boldsymbol{\varepsilon} = 1\% \begin{bmatrix} 0.1060 \\ -0.0241 \\ 0.0144 \\ -0.0633 \\ -0.1804 \\ 0.0971 \end{bmatrix}_{X_1-X_2-X_3}$ | $\boldsymbol{\sigma} = \frac{0.01k^{ab}}{l} \begin{bmatrix} 0.6721 \\ 0.0754 \\ 0.2525 \\ -0.1379 \\ -0.4192 \\ 0.2290 \end{bmatrix}_{X_1-X_2-X_3}$ | $\boldsymbol{\varepsilon} = 1\% \begin{bmatrix} 0.1763 \\ -0.0419 \\ -0.0381 \\ -0.0022 \\ 0 \\ 0 \end{bmatrix}_{\alpha=0.5, \beta=0.7, \gamma=0.0}$ | $\boldsymbol{\sigma} = \frac{0.01k^{ab}}{l} \begin{bmatrix} 1 \\ 0 \\ 0 \\ 0 \\ 0 \\ 0 \end{bmatrix}_{\alpha=0.5, \beta=0.7, \gamma=0.0}$ |
| $\alpha = 0.8$ $\beta = 0.3$ $\gamma = 0.0$ | $\boldsymbol{\varepsilon} = 1\% \begin{bmatrix} 0.1351 \\ -0.0197 \\ -0.0197 \\ -0.0468 \\ -0.1294 \\ 0.1342 \end{bmatrix}_{X_1-X_2-X_3}$ | $\boldsymbol{\sigma} = \frac{0.01k^{ab}}{l} \begin{bmatrix} 0.8075 \\ 0.0991 \\ 0.0935 \\ -0.0962 \\ -0.3038 \\ 0.3128 \end{bmatrix}_{X_1-X_2-X_3}$ | $\boldsymbol{\varepsilon} = 1\% \begin{bmatrix} 0.1763 \\ -0.0431 \\ -0.0374 \\ -0.0014 \\ 0 \\ 0 \end{bmatrix}_{\alpha=0.8, \beta=0.3, \gamma=0.0}$ | $\boldsymbol{\sigma} = \frac{0.01k^{ab}}{l} \begin{bmatrix} 1 \\ 0 \\ 0 \\ 0 \\ 0 \\ 0 \end{bmatrix}_{\alpha=0.8, \beta=0.3, \gamma=0.0}$ |
| $\alpha = 1.2$ $\beta = 1.5$ $\gamma = 0.0$ | $\boldsymbol{\varepsilon} = 1\% \begin{bmatrix} -0.0382 \\ 0.1464 \\ -0.0143 \\ -0.1469 \\ -0.0150 \\ 0.0315 \end{bmatrix}_{X_1-X_2-X_3}$ | $\boldsymbol{\sigma} = \frac{0.01k^{ab}}{l} \begin{bmatrix} 0.0064 \\ 0.8631 \\ 0.1304 \\ -0.3356 \\ -0.0292 \\ 0.0752 \end{bmatrix}_{X_1-X_2-X_3}$ | $\boldsymbol{\varepsilon} = 1\% \begin{bmatrix} 0.1763 \\ -0.0429 \\ -0.0395 \\ -0.0026 \\ 0 \\ 0 \end{bmatrix}_{\alpha=1.2, \beta=1.5, \gamma=0.0}$ | $\boldsymbol{\sigma} = \frac{0.01k^{ab}}{l} \begin{bmatrix} 1 \\ 0 \\ 0 \\ 0 \\ 0 \\ 0 \end{bmatrix}_{\alpha=1.2, \beta=1.5, \gamma=0.0}$ |